\documentclass[12pt]{iopart}

\usepackage{graphicx}
\usepackage{bm}
\usepackage[serbian,english]{babel}
\usepackage{color}

\usepackage{lipsum}
\usepackage{acronym}
\usepackage{subfig}
\usepackage{placeins}

\begin{document}

\title[Periodic three-body orbits with vanishing angular momentum] 
{Periodic three-body orbits with vanishing angular momentum in 
the Jacobi-Poincar\' e ``strong'' potential}

\author{V. Dmitra\v sinovi\' c$^{1}$, Luka V. Petrovi{\' c}$^2$ and Milovan \v Suvakov$^1$}

\address{$^1$ Institute of Physic Belgrade, Belgrade University, Pregrevica 118, 
Zemun, P.O.Box 57, 11080 Beograd, Serbia \\
$^2$ Faculty of Physics, University of Vienna, 
Boltzmanngasse 5, 1090 Vienna, Austria \\
present address: ETH Z\"urich, Otto-Stern-Weg 1, 8093 Z\"urich, Switzerland}
\ead{dmitrasin@ipb.ac.rs,luka.petrovic@gmail.com,suki@ipb.ac.rs}
\vspace{10pt}
\begin{indented}
\item[]August 2017
\end{indented}

\begin{abstract}
Moore \cite{Moore1993} and Montgomery \cite{Montgomery2004} have argued that planar
periodic orbits of three bodies moving in the Jacobi-Poincar\' e, or the ``strong'' pairwise
potential $\sum_{i>j}\frac{-1}{r_{ij}^2}$, can have all possible topologies.
Here we search systematically for such orbits with vanishing angular momentum
and find 24 topologically distinct orbits, 22 of which are new, in a small section of the 
allowed phase space, with a tendency to overcrowd, due to overlapping initial conditions. 
The topologies of these 24 orbits belong to three 
algebraic sequences defined as functions of integer $n=0,1,2, \ldots$. Each sequence extends
to $n \to \infty$, but the separation of initial conditions for orbits with $n \geq 10$ becomes
practically impossible with a numerical precision of 16 decimal places. 
Nevertheless, even with a precision of 16 decimals, it is clear that in each sequence both 
the orbit's initial angle $\phi_n$ and its period $T_n$ approach finite values in
the asymptotic limit ($n \to \infty$). Two of three sequences are overlapping in the sense that 
their initial angles $\phi$ occupy the same segment on the circle and their asymptotic values 
$\phi_{\infty}$ are (very) close to each other. The actions of these orbits rise linearly with 
the index $n$ that describes the orbit's topology, which is in agreement with the Newtonian case. 
We show that this behaviour is consistent with the assumption of analyticity of the action as a function of period.
\end{abstract}

%
\vspace{2pc}
\submitto{\JPA}
%
%
%

\section{Introduction}
\label{s:Introduction}

One singular homogeneous potential stands out for its properties: the attractive 
pairwise $\sum_{i<j}-\frac{G m_i m_j}{r_{ij}^2}$,  also known as the Jacobi-Poincar\' e, or 
the ``strong'' potential has long been known to lead to a simplification of the classical 
N-body problem due to its dilation/conformal symmetry.
In the 19th century Jacobi discovered its dilational symmetry \cite{Jacobi1843} and 
Poincar\'e \cite{Poincare:1896} discussed some properties of its action integral. 
More recently \cite{Calogero:1969xj,Khandekar:1972,Moser1975} 
it was shown that in one spatial dimension, this potential leads to integrable dynamics 
of the celebrated Calogero-Moser model.

In the meantime, this potential has acquired a physical significance:
1) it is part and parcel of post-Newtonian gravity, responsible for the gravitational 
collapse \cite{Poisson:2014}; and
2) it is central to Efimov three-body bound states in molecular physics, see 
Refs. \cite{Bloom:2014,Barth:2015} and references therein.
Thus, it appears important that the classical dynamics in this potential be studied in
more dimensions than one.

In two spatial dimensions the classical dynamics of this potential has been studied only
sporadically, mostly emphasizing mathematical aspects, see Refs. 
\cite{Montgomery2004,Fujiwara2004a,Tosel2000}, with only one significant numerical
study \cite{Moore1993}. It is known that this potential leads to a reduction of the 
number of degrees of freedom by unity, but not to integrability of the system, at least
for unequal masses, \cite{Tosel2000}. That reduction implies a significant simplification, 
at least for periodic orbits.

The authors of Refs. \cite{Moore1993,Montgomery2004} have argued, along different lines,
that periodic three-body orbits with all possible topologies, which is a 
countably infinite set, must exist in this potential; that is not the case for the 
Newtonian potential. Here we note that before the present paper, Moore \cite{Moore1993} had 
found two numerical periodic three-body orbits in the strong potential, one of which (the 
figure-eight) exists in the Newtonian potential and was rediscovered in Ref. \cite{Fujiwara2004a}. 
Moore's second periodic orbit (figure-eight on the shape sphere, see Sect. \ref{ss:Results}) 
in the strong potential does not exist in Newtonian gravity, however.
Thus, a plethora of 
previously unknown types of periodic orbits appears to be beckoning. 
Recent numerical searches for periodic orbits in the Newtonian potential have produced
more than 200 new topologically distinct periodic orbits, see  
Refs. \cite{Suvakov:2013,Suvakov:2013b,Shibayama:2015,Martynova2009,Iasko2014,Suvakov:2014,Dmitrasinovic:2016,Jankovic:2015}. 
The next natural step is to apply the same methods to the three-body problem in the 
strong potential. 

Now, mathematical existence theorems do not necessarily tell one where to search and find the 
respective solutions, merely that they exist. Fortunately, in the case of the strong potential
the phase space of initial conditions 
has fewer dimensions than the one in the case of Newtonian potential. Moreover, it has been shown that
every periodic orbit (with one exception, that has non-zero angular momentum, however) must 
pass through a collinear configuration at least once during its period, 
\cite{Montgomery2002,Fujiwara2004a}, which further reduces the space of initial configurations. 
These facts suggest that the numerical search for periodic orbits in the strong potential should 
be simpler than in the Newtonian one. 

Here we report the results of our search: we found 24 periodic solutions, 22 of which are 
new,\footnote{The two orbits previously discovered by Moore, Ref. \cite{Moore1993}, 
are the figure-8 orbit in configuration space, with topology ${\tt abAB} $ and the
figure-8 in the shape space, with topology ${\tt ab}$.} 
falling into three algebraically defined (infinite) sequences. 
Many other candidates for periodic orbits have been detected, but could not be clearly separated
within our self-imposed numerical precision requirements, for reasons that will be explained 
shortly. 

The obtained results are interesting for the following reasons:
(i) the distribution of periodic orbits in the (phase) sub-space of initial conditions is 
far from uniform: there are vast regions that are entirely devoid of periodic orbits, whereas
several small regions are severely overcrowded with (infinitely) many closely packed - indeed 
overlapping - orbits; 
(ii) all periodic orbits discovered thus far belong to three sequences, indexed by 
integer $n=0,1, \ldots$; each sequence having an 
accumulation point of its own, both as a function of the initial 
angle $\phi_0$ and as a function of the period $T$;
(iii) the sensitivity to ``errors'', i.e. to differences between the fiducial and the actual 
values of the initial conditions, rapidly increases with the sequential index $n$;
(iv) the (almost) linear dependence of the scale-invariant period on the
sequential index $n$, or equivalently on topology, that was observed in the Newtonian case 
\cite{Dmitrasinovic:2015}, breaks down here spectacularly - the periods do not increase 
linearly, but approach finite asymptotic accumulation points (a different one for each sequence);
(v) a linear relation between the initial conditions - angles - and the periods holds
in each sequence separately in the asymptotic limit $n \to \infty$;
(vi) a linear relation between the (scale-invariant) action and the sequential 
index $n$, or equivalently the topology, holds in the asymptotic limit $n \to \infty$. \footnote{In 
other homogeneous potentials one linear dependence would imply the other, but not here, see 
Sect. \ref{ss:action}.} 

We are not aware of another system of ordinary differential equations with such, or similar 
distribution of periodic solutions. At least one (vi) property of these solutions, 
the asymptotically linear relation, as well as the existence of small non-linear 
corrections that disappear in the $n \to \infty$ limit, can be explained in terms of analytic 
properties of the action.

This paper consists of six sections: after some preliminaries in Sect. \ref{s:Preliminaries}, 
wherein we remind the readers
about the basic facts of periodic orbits in the Jacobi-Poincar{\' e} ``strong'' 
$\frac{-1}{r^2}$ potential must have zero total energy and constant hyper-radius, 
we present our search method in Sect. \ref{s:Search}. Then, in Sect. \ref{ss:Results} we show 
our results - the 22 new orbits that together with two previously known orbits form three sequences.
In Sect. \ref{s:Regularities} we display the action ``quantization'' regularities within the three 
sequences. In Sect. \ref{ss:complex} we show some mathematical argument based on analyticity
of the obtained solutions.
Finally, in Sect. \ref{s:summary} we summarize and draw our conclusions. 
\ref{s:Numerical} contains a description of the return proximity function's dependence on 
the numerical accuracy of initial conditions. 
\ref{s:Calogero} contains 
detailed calculation of the action of the periodic three-body orbit in the (1D) Calogero-Moser
model. In 
\ref{s:analytic_evid} we discuss the analytic properties of the action 
expressed in terms of complex variables.



\section{Search for periodic solutions}
\label{s:Search}

Two-dimensional motions in the strong potential can be divided into (at least) three types:
1) positive energy ($E > 0$) leads either to infinite expansion (for positive values of the 
time derivative of the hyper-radius, ${\dot R} > 0$), or to collapse of the system  (for 
negative values of the time of the hyper-radius, ${\dot R} < 0$);
2) negative energy ($E < 0$) always leads to the collapse of the system; and
3) zero energy ($E = 0$) motions with: (a) non-zero values of the 
time derivative of the hyper-radius, ${\dot R} \neq 0$) lead to either collapse or infinite
expansion, whereas (b) only vanishing time derivative of the hyper-radius, ${\dot R} \neq 0$) 
leads to dilation-invariant motion(s), which includes, but is not limited to, periodic ones.
Thus periodic orbits occupy a small subset of all possible motions. Further, vanishing
angular momentum condition ensures that the corresponding three-body orbit must be in a plane. 

Thus, in case (3.b), the motions of the system can be described by two dimensionless degrees of 
freedom, such as two angles parametrizing a unit-radius sphere $S_2$ in three-dimensional 
``shape space''. There are 
three points on this sphere that correspond to two-body collisions, and thus to infinite kinetic 
and potential energies; these three points determine the topology of a periodic solution. 
The ``volume'' of such a configuration space (i.e. the area of the sphere) is finite, and the 
corresponding phase space (excluding the collision points) is also compact, which makes the 
subspace of periodic orbits finite. Moreover, due to the at least one syzygy 
theorem \cite{Montgomery2002}, the initial configuration space may be further reduced 
to just the equator on the shape sphere, excluding the three collision points.

\subsection{Preliminaries}
\label{s:Preliminaries}

The planar, or two-dimensional (2D) three-body motions in the $-\frac{1}{r^2}$ potential display
several peculiarities that we list below.

1) All periodic solutions in the strong potential three-body problem
must have exactly zero energy and the hyper-radius (the ``scalar moment of inertia'') 
must stay constant at all times. These facts follow from the Lagrange-Jacobi identity, 
or the virial theorem. 

2) We use the fact, shown in Ref. \cite{Montgomery2002} for the Newtonian potential, and 
in Ref. \cite{Fujiwara2004a} for the strong potential, that all but one periodic orbits must 
cross the equator on the shape sphere at least once during their periods (``the syzygy theorem'').

These two facts reduce the dimensionality of the initial state phase space and 
allow us to use a point on the equator of the shape sphere as an initial configuration.
This facilitates the search for periodic solutions. Moreover, the virial theorem 
also implies that the action of a periodic orbit is {\it not} proportional
to the (vanishing) energy of the periodic solution.

``Normalizing'' orbital periods by spatially scaling solutions to the same energy level 
is not possible in this potential, due to the vanishing energy. 
Consequently, the constant appearing in Kepler’s third law - the ``scale-invariant period'' 
$T|E|$ is trivial here, as it always equals zero. Instead, one keeps the scale, or the 
hyperradius of all orbits invariant at one fixed value. The periods depend on this scale,
as $T \sim R^2$, i.e., the product $T R^{-2}$ is scale invariant.
The minimized action $S_{\rm min}$ of periodic orbits is scale invariant, however,
and can be used instead of the above ``scale-invariant period'' $T|E|$, to 
``measure'' the topological dependence of orbits 
in the case of strong potential, in the sense of Refs. \cite{Dmitrasinovic:2016,Dmitrasinovic:2015}.

\subsection{Initial conditions}
\label{ss:Initial}

As stated above, the two main distinctions of the Jacobi-Poincar{\' e} $\frac{-1}{r^2}$ ``strong'' 
potential are that: 
1) the energy of all periodic orbits must be zero $E=0$; 2) the hyper-radius, or ``overall size'', 
$R = \sqrt{{\bm \rho}^2 + {\bm \lambda}^2} = 
\sqrt{\frac{1}{3} \sum_{i<j}^{3}({\bf r}_{i} - {\bf r}_{j})^2}$
of all periodic orbits must remain constant at all times $R(t)=R(0)$. These two conditions
reduce the number of degrees of freedom of the planar equal-mass three-body problem from 
three to two. 
The conditions of 1) vanishing angular momentum, and 2) equal masses of all three bodies, 
are our own choice, in this paper.

There are $12 = 3 \times 4$ independent kinematic variables that define the 
state vector (in phase space)
${\bf X}(t)=\left ({\bf r}_{1}(t), {\bf r}_{2}(t), {\bf r}_{3}(t),{\bf p}_{1}(t),{\bf p}_{2}(t),{\bf p}_{3}(t)\right)$ 
of a three-body system (in two dimensions):  
for each of three bodies there are two coordinates ($x$ and $y$) for each body, and 
two components (${\bf p}_{x} = v_x$ and ${\bf p}_{y}= v_y$) of its velocity vector. 
We set $G=1=m_1=m_2=m_3$, which does not reduce the generality of our results, as 
certain scaling rules hold, see Ref. \cite{Landau}. 
The scaling rules allow one to obtain solutions for any (real, positive)
value of $G$ and/or of the common mass $m_1=m_2=m_3$ from the solutions presented here.
Of course, distinct-mass orbits cannot be obtained from the equal-mass limit ones.
The choice of center-of-mass system (${\bf v}_1 + {\bf v}_2 + {\bf v}_3 =0$) cuts down 
this number to eight independent variables, 
as there are (only) two independent relative coordinate vectors and two corresponding velocities.
The choice of vanishing angular momentum $(L=0)$ reduces this number down to seven.

We choose the so-called Euler initial configuration, the three bodies being collinear, say on the x-axis, 
with the distance between the bodies number ``one'' and number ``two'' equaling two units (2), and with the 
body number ``three'' at the mid-point between bodies number ``one'' and number ``two''. 
That sets the (initial state) hyper-radius at $R=\sqrt{2}$. As the hyper-radius must remain 
constant during periodic motions, we are left with six independent variables. The conditions 
$L=0$, ${\dot R} =0$, and ${\bf v}_1 + {\bf v}_2 + {\bf v}_3 =0$ put together imply 
\[{\bf v}_1 = {\bf v}_2= −\frac12 {\bf v}_3,\]
thus inferring that only one velocity two-vector is independent in this choice of initial 
conditions.

Finally, demanding ${\dot R} =0$, leaves the system with two independent variables: the angle $\phi$ 
between the $x$-axis and the velocity 2-vector $v_1$, and the overall size $R$. Due to the zero-energy 
condition $E=0$ the size $R$ of the system, which has already been set at $R = \sqrt{2}$ 
by our choice of initial positions, determines the value
of the initial kinetic energy $T$ as $T = -V(R)$, thus leaving the angle $\phi$ as the only 
free variable.

This means that, in order to find periodic orbits passing through the Euler point, the only 
variable that 
can be varied in this sub-space of initial conditions is the angle $\phi$ between the two components of 
the vector ${\bf v}_1= \left(v_{x1}, v_{y1} \right)$: $\tan \phi  = \frac{v_{y1}}{v_{x1}}$.

\subsection{Search Method}
\label{ss:Method}

The return proximity function $d({\bf X}_0,T_0) = d(\phi_0,T_0)$ in phase space 
is defined as the absolute minimum of the distance from the initial condition by
$d({\bf X}_0,T_0)=\min_{t \le T_0} \vert {\bf X}(t)-{\bf X}_0 \vert$,
where
\begin{equation}
\left \vert {\bf X}(t)-{\bf X}_0 \right \vert =
\sqrt{\sum_i^3 [{\bf r}_{i}(t) - {\bf r}_{i}(0)]^2 +
\sum_i^3 [{\bf p}_{i}(t) - {\bf p}_{i}(0)]^2 }
\end{equation}
is the distance (Euclidean norm) between two 12-vectors in phase space
(the Cartesian coordinates and velocities of all three bodies without
removing the center-of-mass motion).
We define the return time $\tau({\bf X}_0,T_0)$ as the time for which
this minimum is reached. 
Numerical minimization of this function has been used in several successful 
searches for periodic orbits in the Newtonian potential 
\cite{Suvakov:2013,Suvakov:2013b,Shibayama:2015,Martynova2009,Iasko2014,Suvakov:2014,Dmitrasinovic:2016,Jankovic:2015}.
Those searches were conducted in a two-dimensional subspace, because the hyper-radius $R$
is variable in the Newtonian three-body periodic orbits. The choice 
of the initial conditions has been explained in Sect. \ref{ss:Initial}.

In order to look for periodic solutions numerically, we have discretized
the search window in the one-dimensional subspace  
and calculated the return proximity function $d(\phi = \phi_0,T_0)$ for each grid 
point up to some pre-defined upper limit on the integration time $T_0$, which was 
set at $T_0 = 10$. 
We shall see that the period $T$ does not grow with the 
length of the orbit's ``word'', but rather $T$ approaches an asymptotic limit
that depends on the algebraic structure of the ``word''.

\subsection{Solving the equations of motion}
\label{s:Solving}

We did the calculation in two stages: at first we used a fourth-order Runge-Kutta-Fehlberg integrator, 
for the full set of three-body equations in Cartesian coordinates, which, of course, does not conserve 
the energy, or the hyper-radius. 
When that integrator started showing its limitations we wrote a new integrator 
for the two true dynamical variables (two hyperangles on the shape sphere), thus eliminating 
two  constants of motion (the hyper-radius and the angular momentum) from the start.
The third constant of motion is the energy $E$, which must vanish ($E=0$) for all periodic
orbits; that constraint can be ``hard-wired'' into the code using the Hamiltonian formalism 
that is manifestly symplectic and leads to a (much) higher accuracy even with a 
Runge-Kutta algorithm. 
That allowed us to determine the accuracy of the previous RKF calculation, as shown in 
detail in \ref{s:Numerical}.

It is well known that the planar three-body dynamics can be expressed in terms of 
hyper-spherical coordinates \cite{Suvakov:2010} as a function of 
the hyper-radius $R$ and the shape-sphere unit vector:
\begin{equation}
{\hat {\vec n}} = \left(n_x^{'},n_y^{'},n_z^{'} \right) = 
\left(\frac{2 {\bm \rho} \cdot {\bm \lambda}}{R^2}, \frac{{\bm \lambda}^2 - {\bm \rho}^2}{R^2}, 
\frac{2 ({\bm \rho} \times {\bm \lambda}) \cdot {\bf e}_z}{R^2}  \right).
\end{equation}
As stated already in Sect. \ref{ss:Initial}, the hyper-radius $R$ is a constant of motion for 
all periodic solutions in this potential. We can assume without loss of generality that $R=1$. 
Therefore, the configuration space is two dimensional in this case. 
We choose the polar angle $\alpha \in [0,\pi]$ and the azimuthal angle $\beta \in [0,2\pi]$ as
the dynamical variables on the shape-sphere.  
The phase space is four dimensional with these two angles as the generalized coordinates and 
their two conjugate generalized momenta.  
Therefore, the equations of motion are four first-order Hamilton's equations.

One can implement the zero-energy constraint $E=0$ into these four equations and thus 
eliminate one variable and one equation. 
The initial-state variables defined in Sect. \ref{ss:Initial} can be written as 
functions of a single parameter $\phi$ which was defined as the angle between the two components of 
the velocity vector ${\bf v}_1= \left(v_{x1}, v_{y1} \right)$: $\tan \phi  = \frac{v_{y1}}{v_{x1}}$,
or equivalently as the angle of the angular velocity, measured with respect to the equator, on 
the shape sphere:
\begin{equation}
\dot{\alpha}=\sqrt{8 |V(\alpha,\beta)|} \sin \phi,
\end{equation}
\begin{equation}
\dot{\beta}=\sqrt{8 |V(\alpha,\beta)|} \frac{\cos \phi}{\sin \alpha},
\end{equation}
where 
\begin{equation}
V=V(\alpha,\beta)=-\sum_{l=0}^{2} \frac{1}{1 - \sin \alpha \cos (\beta + 2 \pi l /3)}
\end{equation}
is the hyper-angular potential as a function of $\alpha$ and $\beta$. 
One can easily check that this change 
of variables 
identically satisfies $E=0$, by calculating 
\begin{equation}
E = T(\dot{\alpha},\dot{\beta},\alpha,\beta) + V(\alpha,\beta)=0,
\end{equation}
where the kinetic energy $T$ is given by the formula:
\begin{equation}
T(\dot{\alpha},\dot{\beta},\alpha,\beta)=\frac{1}{8} \left(\dot{\alpha}^2 + \dot{\beta}^2 \sin^2 \alpha \right).
\end{equation}

Three variables $\alpha$, $\beta$, and $\phi$ completely define the state of the system. 
Using this set of variables we now have three first-order differential equations of motion, 
instead of four; they are:
\begin{eqnarray}
\dot{\alpha} &=& \sqrt{8 |V(\alpha,\beta)|} \sin \phi, \\
\dot{\beta} &=& \sqrt{8 |V(\alpha,\beta)|} \frac{\cos \phi}{\sin \alpha}, \\
\dot{\phi} &=& \sqrt{8 |V(\alpha,\beta)|} \frac{\cos \alpha}{\sin \alpha} \cos \phi + 
\sqrt{\frac{2}{|V|}} A \cos \phi + \sqrt{\frac{2}{|V|}} B \frac{\sin \phi}{\sin{\alpha}},
\end{eqnarray}
where $A$ and $B$ are:
\begin{eqnarray}
A  &=&  \frac{\partial |V(\alpha,\beta)|}{\partial \alpha}=
\sum_{l=0}^{2} 
\frac{\cos \alpha \cos (\beta + 2\pi l /3)}{\left(1-\sin \alpha \cos (\beta + 2\pi l /3)\right)^2},\\
B &=& -\frac{\partial |V(\alpha,\beta)|}{\partial \beta}=
\sum_{l=0}^{2} \frac{\sin \alpha \sin (\beta + 2\pi l /3)}{\left(1 - \sin \alpha \cos (\beta + 2\pi l/3)\right)^2}.
\end{eqnarray}
The initial conditions defined previously in Sect. \ref{ss:Initial} correspond to $\alpha=\pi/2$,
$\beta = \pi$, with $\phi$ taken as a free parameter.
Using the variables $\alpha$, $\beta$, and $\phi$, the return proximity function
can be redefined as follows
\begin{equation}
d(\alpha_0,\beta_0,\phi_0,T_0)=\sqrt{(\alpha(t)-\alpha(0))^2+(\beta(t)-\beta(0))^2 + (\phi(t)-\phi(0))^2}.
\end{equation}
We solved the equations of motion using an explicit fifth-order Runge-Kutta algorithm \cite{Hairer1993}
implemented in python library SciPy \cite{Python} with best relative tolerance of $10^{-20}$. 
We note that the Hamiltonian 
(first-order) nature of the equations of motion ensures exact conservation of energy $E=0$ at each 
step of the calculation, regardless of the particular algorithm used for the numerics.

The return proximity function has been calculated using linear interpolation between numerically obtained
values of $\alpha(t)$, $\beta(t)$, and $\phi(t)$. This interpolation method creates additional
errors that cannot be explicitly calculated, but can be numerically investigated as in \ref{s:Numerical}.

\subsection{Topological classification of orbits in the shape space}
\label{s:classification}

Any newly found periodic three-body orbit must be identified and classified
so as to be distinguished from previously discovered orbits. For that purpose
we use Montgomery's topological classification \cite{Montgomery1998}:
he noticed the connection between the 
``fundamental group of a two-sphere with three punctures'', i.e. 
the ``free group on two letters'' ${\tt (a,b)}$, and the conjugacy classes 
of the ``projective coloured/pure braid group'' of three strands $PB_3$. 
The utility of this classification becomes apparent when we identify the 
``two-sphere with three punctures'' with the 
shape-space sphere and the three two-body collision points with the punctures. 

Graphically, this method amounts to classifying closed curves according to 
their ``topologies'' on a sphere with three punctures. A stereographic projection of
this sphere onto a plane, using one of the punctures as the ``north pole'' 
effectively removes that puncture to infinity, and reduces the problem to one 
of classifying closed curves in a plane with two punctures. That leads to the 
aforementioned free group on two letters ${\tt (a,b)}$, where (for definiteness) 
${\tt a}$ denotes a clockwise ``full turn'' around the right-hand-side puncture, 
and ${\tt b}$ denotes the counter-clockwise full turn around the other puncture, see Ref. 
\cite{Suvakov:2013}. 
For better legibility we denote their inverses by capitalized letters ${\tt a^{-1}=A}$, 
${\tt b^{-1}=B}$. 

Of course, there need not be only one solution with a particular topology:
indeed orbits with different values of the angular momentum and identical topology 
define one (continuous) topological family of orbits. In the Newtonian potential there 
are sometimes multiple orbits\footnote{And sometimes none.} with identical topologies 
and angular momenta, but, as we shall see, that does not happen here.

A specific sequence of letters, or a word, is not
the only possible description of a periodic orbit, because there is no
preferred initial point on a periodic orbit; 
therefore any other word that can be obtained by a cyclic permutation of the letters 
in the original word is an equally good description of such an orbit.
For example the conjugacy class of the free group element 
${\tt aB}$ contains also the element ${\tt A(aB)a = Ba}$. 
The set of all cyclically permuted words is the aforementioned conjugacy class
of a free group element (word). 
Thus, each family of orbits is associated with the conjugacy class of 
a free group element. 

Moreover, the time-reversed orbits correspond to physically identical
solutions, but their free group elements and their conjugacy classes are generally
different. So, for example, families of orbits described by {\tt a} and {\tt A} are equivalent,
but families {\tt ab} and {\tt AB} are not because the inverse of {\tt ab} is {\tt BA},
not {\tt AB}.

There is finally one last ambiguity concerning all non-cyclically permutation symmetric orbits
\cite{Dmitrasinovic:2015}. 
One may apply the stereographic projection of the shape sphere onto a plane using any one 
of the three two-body collision points as the North Pole, and in that way one may obtain three 
(at least in principle) different words. When the orbit is cyclically permutation symmetric, 
this ambiguity disappears, for all other orbits one may switch to the three-symbol
labeling scheme \cite{Dmitrasinovic:2016} to resolve this issue rigorously. But, for all practical 
purposes one need not go 
to such lengths, as it is sufficient to make sure that the same two-body collision point 
(``puncture'') has been used as the North Pole in the stereographic projection for all the orbits
treated. Then, the same algebraic relations must hold among the orbits' words, irrespective of the 
choice of the puncture.

For a working algorithm that ``reads'' an orbit's word, see the Appendix to Ref. \cite{Suvakov:2014}.

\section{Results}
\label{ss:Results}

\begin{figure}
\hskip48pt
\includegraphics[width=0.90\columnwidth,,keepaspectratio]{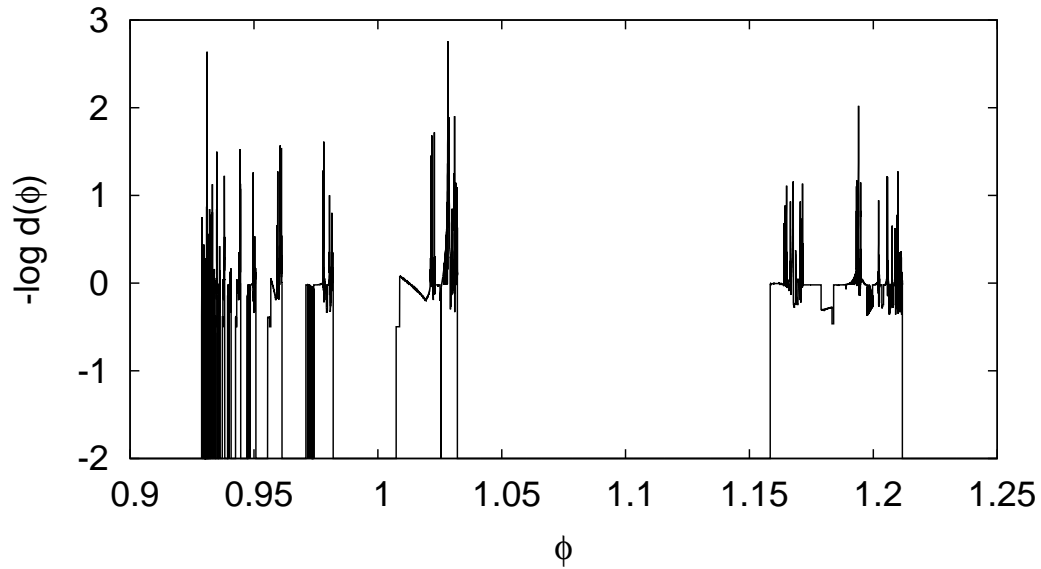}
\vskip48pt
\caption{The negative logarithm of the return proximity function $−\log(d)$ resulting from 
a scan over the initial angle range $\phi \in [0.9, 1.25]$.}
\label{fig:scanpokrugu}
\end{figure}

\begin{figure}
\centering\includegraphics[width=0.75\columnwidth,,keepaspectratio]{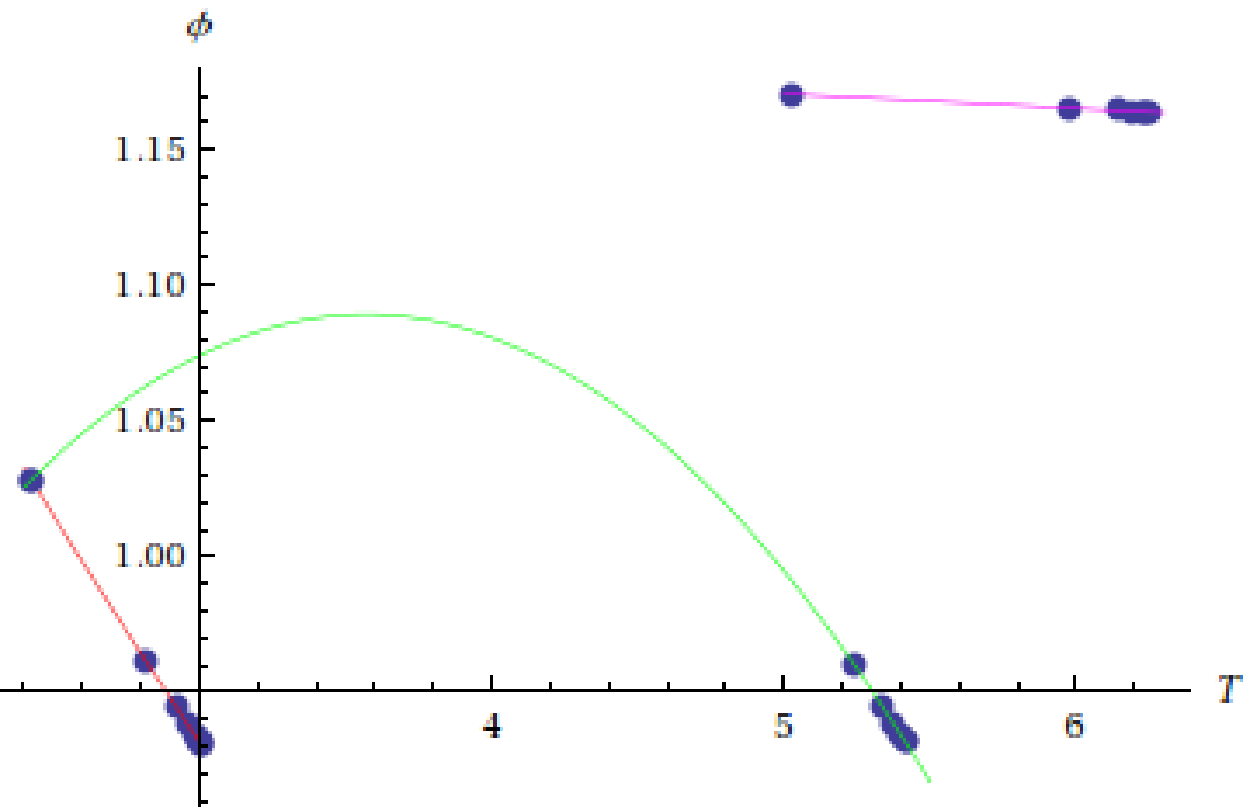}
\caption{(online colour) The initial angle $\phi$ as a function of the period $T$, for 
orbits in the three sequences, Tables \ref{tab:IC1},\ref{tab:IC2},\ref{tab:IC3} together with 
quadratic fits (colored continuous lines).
Three accumulation points are clearly visible, as well as the fact that one sequence
obstructs the ``visibility'' of another.}
\label{fig:Phi_T_total}
\end{figure}

In Fig.\ref{fig:scanpokrugu} 
we show the $-\log(d(\phi_0))$ dependence on the initial angle $\phi = \phi_0$ in the 
range $\phi \in [0.9, 1.25]$. There one can see many overlapping peaking structures 
some of them broad (``hills''), others quite narrow 
(``trees'', or ``spikes''), in roughly four distinct regions (``groves''). 
Several properties of the newly discovered orbits shine through immediately:
\begin{enumerate}
 \item periodic orbits exist only in several 
 finite, small segments of the (quarter-circular) perimeter [in the angular range 
 $0.9 \leq \phi \leq 1.25$ (in radians)], 
 rather than on the whole quarter-circle $0 \leq \phi \leq \pi/2 = 1.507$;

 \item We found around 100 peaking structures (at the present level of precision) in this region, 
 only 24 of which kept increasing their $-\log(d)$ values as the precision was improved, see 
 Appendix \ref{s:Numerical}. These 24 initial conditions appear in three distinct regions 
 (``groves'') in Fig. \ref{fig:scanpokrugu}, and are tabulated in Tables \ref{tab:IC1}, 
 \ref{tab:IC2}, \ref{tab:IC3}. 

 \item The remaining (roughly 80) peaking structures are interspersed through all distinct regions 
 (``groves''). These structures (probably) correspond to (quasi-periodic) orbits that do not pass exactly 
 through the Euler point, but come close to it. Therefore, their  $-\log(d)$ values cannot 
 be increased beyond some limiting value.  Note that one ``grove'', $\phi \in (1.2, 1.22)$, consists entirely 
 of such ``quasi-periodic'' orbits. The corresponding periodic orbits must be searched for in a 
 two-parameter space (angle $\phi$ and the ``displacement'' from the Euler point), which is beyond 
 our present scope.

 \item The periodic orbits found thus far can be classified into three sequences with well-defined 
 algebraic structures describing the topologies of the orbits, \cite{Montgomery1998}, very much as 
 in the Newtonian case, Ref. \cite{Dmitrasinovic:2016}.
 The topologies $w_n^{(i)}$ of solutions belonging to three sequence are defined as 
 $w_n^{(1)} = {\tt a^n b^n}$, $w_n^{(2)} = {\tt A^n B^n a b}$, $w_n^{(3)} = {\tt a^n b^n a b}$, 
 with integer $n=0,1,2,\ldots$.
 
 \item The periods $T_n$ and initial angles $\phi_n$ of the $n$-th periodic orbit 
 converge to one of three accumulation points, depending on the sequence, see 
 Fig. \ref{fig:Phi_T_total}. 
 It follows immediately that there is no linear relation between the ordering number $n$ 
 and the period $T_n$ of the $n$-th orbit in any of the three sequences, in contrast to 
 the Newtonian case.

 \item This accumulation of orbits' initial conditions makes the search for higher-$n$, i.e., 
 beyond some (fairly small number, such as 10) $n$, orbits in the same sequence very difficult,
 if not completely impossible. This is because for some value of the index $n$ the difference 
 $\Delta \phi_n = |\phi_n - \phi_{n-1}|$ between two sequential initial conditions becomes 
 comparable with, or smaller than the numerical accuracy, 
 see 
 \ref{s:Numerical}. It should therefore be clear that finding 
 further sequences with longer periods will be severly limited by the available numerical accuracy. 
 Thus, at the present time we can only conjecture that initial conditions of orbits in sequences 
 with topologies described by 
 ${\tt a^k b^k a^n b^n}$, $n= k,k+1,k+2, \cdots$ and $k = 2,3,, \cdots$, are ``hidden behind'' 
 the two already observed ones with $k=0,1$. 

 \item Not one topological-power (with topology of the form $w^k$, $k=2,3,\cdots$) orbit has been 
 found, thus far, in this potential, in contrast with the Newtonian potential. If that were a 
 general rule, it 
 would explain the absence of the $n=1$ term from the ${\tt a^n b^n a b}$, $n=0,2,3, \cdots$ 
 sequence in Table \ref{tab:IC3}, as this term would be the second power of 
 $({\tt ab})^2 = {\tt (a^1 b^1) a b}$. 

 \item Only one orbit has been found for each topology, i.e., there are no multiple periodic orbits 
 with the same topology, again in contrast to the Newtonian three-body problem. Uniqueness of the 
 figure-eight orbit has been proven in Ref. \cite{Montgomery2004}, so the uniqueness of the here 
 observed orbits leads us to conjecture that all periodic orbits in this potential might be 
 unique.

\end{enumerate}

\begin{table}
\caption{\label{tab:IC1} Initial condition (angle $\phi$), and period of three-body orbits in 
the sequence ${\tt a^n b^n}$, $n=0,1,2 \cdots$, in the ``strong force'', with coupling constant $G$ is 
taken as $G=1$ and equal masses as $m_{1,2,3}=1$. The total energy is zero ($E=0$), so the initial
velocities are determined by a single angle $\phi$. We also show the lowest
return proximity value ${\rm min}~d(\phi,T)$, the period $T$ and the free group element (f.g.e.). 
The orbit denoted by 0 is the Calogero-Schubart colliding one.}
\footnotesize
\begin{tabular}{@{}llll}
\br
$\phi$ & ${\rm T}/2$ & ${\rm min}~d(\phi,T)$ & {\rm f.g.e.} \\
\mr
0  & 0.444445 & $10^{-\infty}$ & ${\tt a^0 b^0}$ \\ 
\hline
1.02824692798800732 & 1.21269962942932730 & $8.56 \times 10^{-10}$ & ${\tt a^1 b^1}$ \\ 
0.96097510395224572 & 1.41012664240185859 & $4.86 \times 10^{-8}$ & ${\tt a^2 b^2}$ \\ 
0.94454479748942055 & 1.46028397633386153 & $3.39 \times 10^{-8}$ & ${\tt a^3 b^3}$ \\ 
0.93814716192692216 & 1.48000532746712521 & $4.71 \times 10^{-8}$ & ${\tt a^4 b^4}$ \\ 
0.93500390397660027 & 1.48973256473400473 & $5.02 \times 10^{-8}$ & ${\tt a^5 b^5}$ \\ 
0.93322897503736779 & 1.49523614367717128 & $1.38 \times 10^{-8}$ & ${\tt a^6 b^6}$ \\ 
0.93212904360729043 & 1.49865060080713763 & $3.15 \times 10^{-8}$ & ${\tt a^7 b^7}$ \\ 
0.93140041145929431 & 1.50091405578946335 & $1.24 \times 10^{-7}$ & ${\tt a^8 b^8}$ \\ 
0.93089288086015565 & 1.50249147597823884 & $6.28 \times 10^{-8}$ & ${\tt a^9 b^9}$ \\  
0.93052521736517435 & 1.50363454265253282 & $6.19 \times 10^{-8}$ & ${\tt a^{10} b^{10}}$ \\ 
\br
\end{tabular}\\
\end{table}
\normalsize

\begin{table}
\caption{\label{tab:IC2} Initial condition (angle $\phi$),  
and period of three-body orbits in the sequence ${\tt a b A^n B^n}$, $n=0,1,2 \cdots$ in
the ``strong force'' potential, with the same values of parameters as in Table \ref{tab:IC1}.
We also show the lowest return proximity value ${\rm min}~d(\phi,T)$, the period $T$ and 
the free group element (f.g.e.).}
\footnotesize
\begin{tabular}{@{}llll}
\br
$\phi$ & ${\rm T}/2$ & ${\rm min}~d(\phi,T)$ & {\rm f.g.e.} \\
\mr
1.02824692798800732 & 1.21269962942932730 & $8.56 \times 10^{-10}$ & ${\tt ab A^0 B^0}$ \\ 
\hline
1.17049635743816727 & 2.51487201239182445 & $3.77 \times 10^{-8}$ & ${\tt a b A^1 B^1}$ \\ 
1.16542247378727315 & 2.98924613111245119 & $4.69 \times 10^{-8}$ & ${\tt a b A^2 B^2}$ \\ 
1.16451391284537076 & 3.07222781280329249 & $3.8 \times 10^{-8}$  & ${\tt a b A^3 B^3}$ \\ 
1.16420804317607640 & 3.10000877826202093 & $5.15 \times 10^{-8}$ & ${\tt a b A^4 B^4}$ \\
1.16406923772482696 & 3.11258622399824114 & $3.8 \times 10^{-7}$  & ${\tt a b A^5 B^5}$ \\
1.16399470076300249 & 3.11933231322032567 & $2.81 \times 10^{-6}$ & ${\tt a b A^6 B^6}$ \\
1.16395009046894526 & 3.12336590267499581 & $4.76 \times 10^{-6}$ & ${\tt a b A^7 B^7}$ \\
\br
\end{tabular}\\
\end{table}
\normalsize

\begin{table}
\caption{\label{tab:IC3} Initial condition (angle $\phi$),  
and period of three-body orbits in the sequence ${\tt a^n b^n a b}$, $n=0,2,3, \cdots$, 
in the ``strong force'' potential, with the same values of parameters as in Table \ref{tab:IC1}.
We also show the lowest return proximity value ${\rm min}~d(\phi,T)$,  
the period $T$ and the free group element (f.g.e.). Note that the $n=1$ orbit is missing.}
\footnotesize
\begin{tabular}{@{}llll}
\br
$\phi$ & ${\rm T}/2$ & ${\rm min}~d(\phi,T)$ & {\rm f.g.e.} \\
\mr
1.02824692798800732 & 1.21269962942932730 & $8.56 \times 10^{-10}$  & ${\tt a^0 b^0 ab}$ \\ 
\hline
0.96053567047115329 & 2.61984543973343476 & $6.41 \times 10^{-8}$ & ${\tt a^2 b^2 a b}$ \\ 
0.94436142254818101 & 2.66897947579246475 & $3.28 \times 10^{-7}$ & ${\tt a^3 b^3 a b}$ \\ 
0.93805794284507649 & 2.68839038409076103 & $6.42 \times 10^{-7}$ & ${\tt a^4 b^4 ab}$ \\ 
0.93495448623723931 & 2.69799863788414607 & $5.99 \times 10^{-6}$ & ${\tt a^5 b^5 a b}$ \\ 
0.93319891411979261 & 2.70345025077037748 & $2.83 \times 10^{-6}$ & ${\tt a^6 b^6 a b}$ \\ 
0.93210944940321705 & 2.70684099103309794 & $9.63 \times 10^{-6}$ & ${\tt a^7 b^7 a b}$ \\ 
0.93138694948558265 & 2.70908665447520480 & $7.65 \times 10^{-6}$ & ${\tt a^8 b^8 a b}$ \\ 
\mr
\end{tabular}
\end{table}

\begin{figure}
\centering\includegraphics[width=0.45\textwidth,,keepaspectratio]{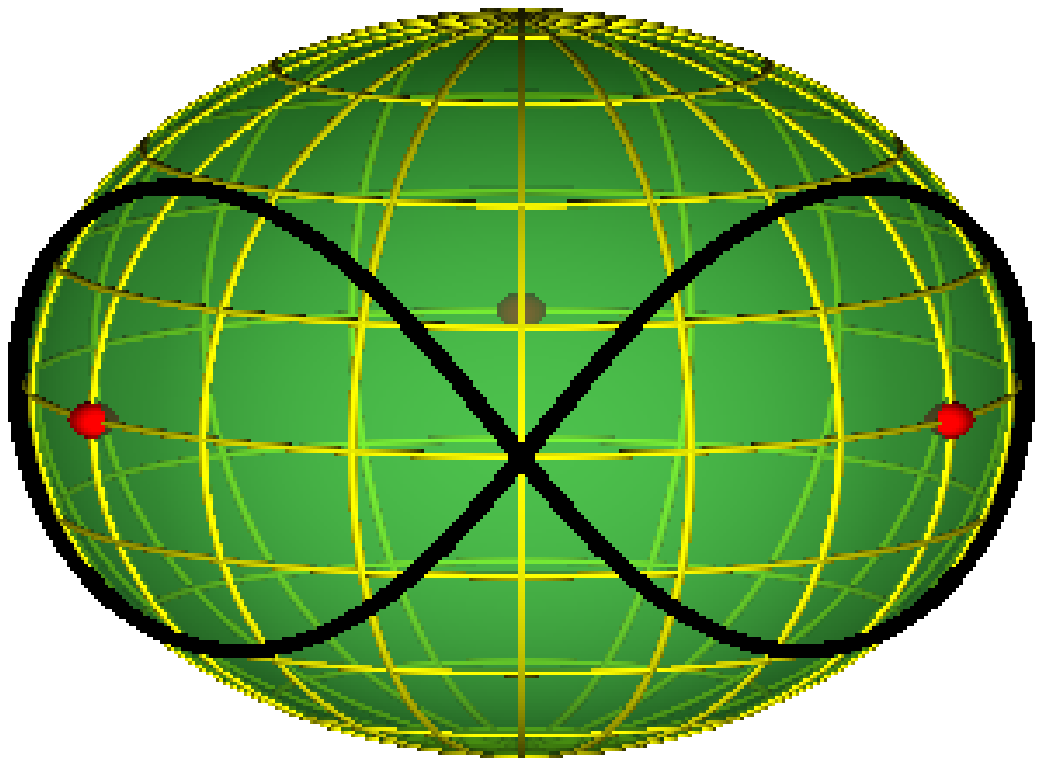},\includegraphics[width=0.45\textwidth,,keepaspectratio]{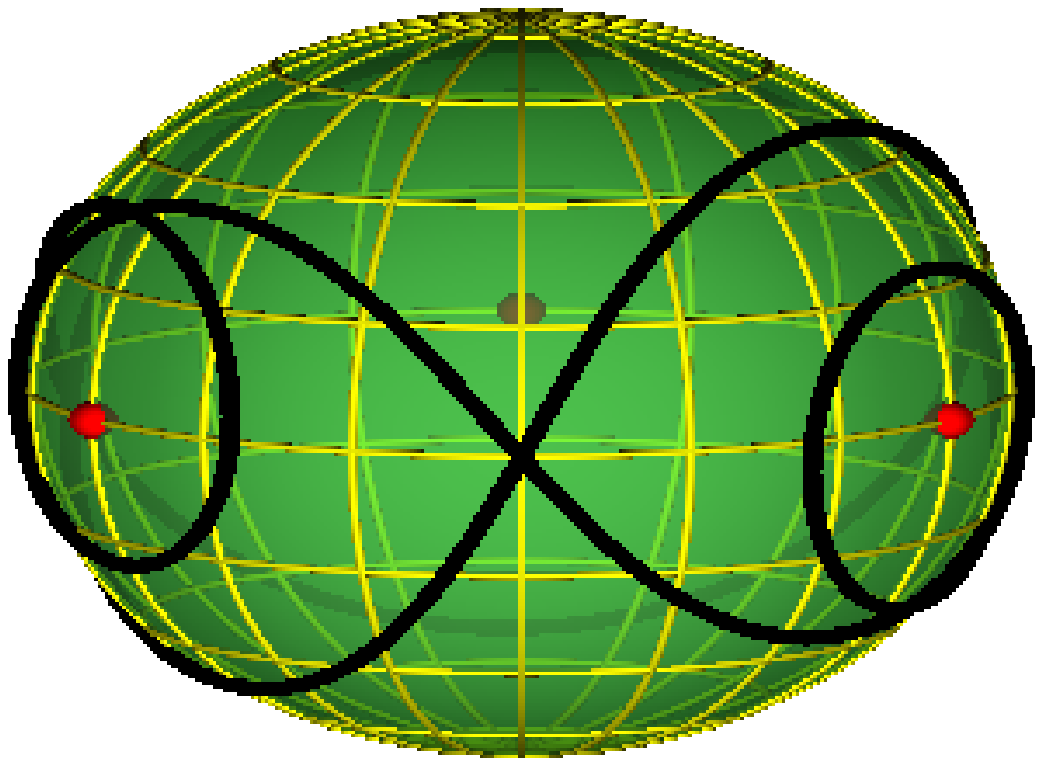}
\caption{(color on line) Orbits on the shape-space sphere: (left hand side) 
``figure-8 on the shape-sphere'' - ${\tt ab}$; (right hand side) orbit ${\tt abA^2B^2}$.}
\label{fig:sphere12}
\end{figure}

\begin{figure}
\centering\includegraphics[width=0.45\textwidth]{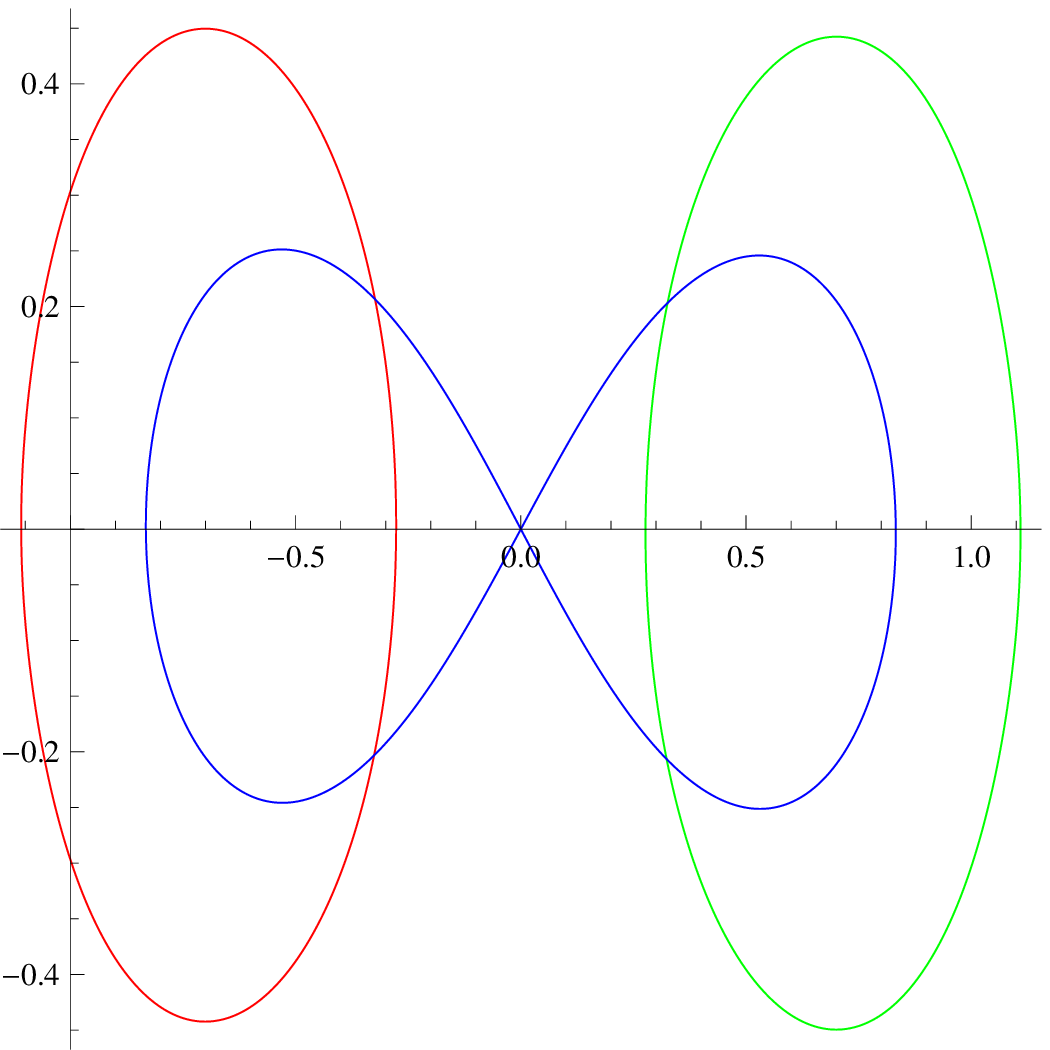},\includegraphics[width=0.45\textwidth]{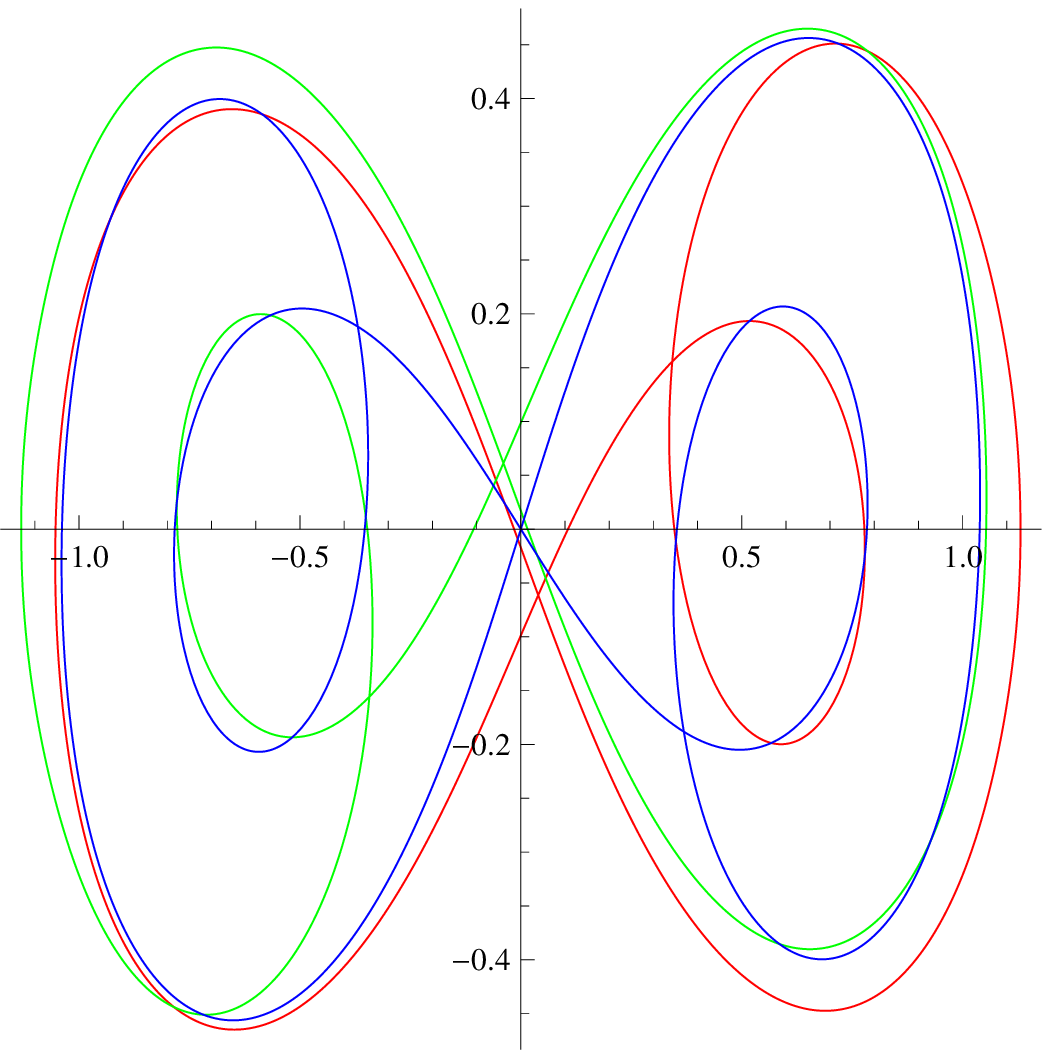}
\caption{(color on line) Orbits in configuration space: (left-hand side) 
``figure-8 on the shape-sphere'' - ${\tt ab}$; (right-hand side)  orbit ${\tt abA^2B^2}$.}
\label{fig:12}
\end{figure}

In Figs. \ref{fig:sphere12} and \ref{fig:12} we show two solutions on the shape sphere and in 
configuration space. These are typical solutions, insofar as they do not approach any of the three 
collision points. Indeed, all the orbits found so far fall between an inner and an outer envelope, 
both in the real and in the shape space: the inner envelope precludes the orbit(s) from getting 
too close to a(ny) collision point. 


\section{Topological dependence of the action}
\label{s:Regularities}

\subsection{Some properties of the action} 
\label{ss:action}

For periodic orbits in a homogeneous potential with (arbitrary) degree of homogeneity $-\alpha$, 
the (minimized) action $S_{\rm min}$ 
can be related to the energy $E$ and period $T$ of the orbit as follows
\[S_{\rm min} = \left(\frac{\alpha + 2}{\alpha - 2} \right) E ~T .\] 
Note that the $\alpha = 2$ case is singular: both the numerator $E$ and the denominator $\alpha - 2$
on the right-hand-side of this identity are zero. 
Consequently, the action $S_{\rm min}$ is not linearly 
proportional to the period $T$, and the two must be separately determined.
Therefore, the periods are not constrained by any ``quantization'' of the action.  

The following approximate linear relation for the minimized actions $S_{\rm min}$, 
of two orbits described by topologies, or free group ``words'', $w$ and $w^k$, 
\[ \frac{S_{\rm min}(w^k)|E(w^k)|^{\alpha/2}}{S_{\rm min}(w)|E(w)|^{\alpha/2}} 
\simeq k = 1,2,3,... \quad \]
has been found numerically in the ($\alpha =1$) Newtonian potential, 
Refs. \cite{Suvakov:2014,Dmitrasinovic:2015,Dmitrasinovic:2016}. 

In an $\alpha = 2$ potential $|E(w^k)| = |E(w)| = 0$ holds, and the (minimized) 
action $S_{\rm min}$ is scale invariant. Then 
\[\frac{S_{\rm min}(w^k)}{S_{\rm min}(w)} \simeq k = 1,2,3,... \quad \]
ought to hold and should be tested; the trouble with this specific example is 
that only one orbit in each topological sector of the
strong potential three-body problem has been found (thus far) - there are no known 
examples of topological powers of any orbit in the $\alpha = 2$ potential.
Therefore, we cannot test this relation directly. 

A more general relation, that the actions 
of all orbits in the same sequence are proportional to the ``length'' $N$ of the word $w$, i.e., 
to the sum of $N = n_a + n_b + n_A + n_B$, was conjectured in Ref. \cite{Dmitrasinovic:2015}.
Thus, one ought to see a correlation between the actions and lengths of orbits' words
\[ \frac{S_{\rm min}(w^{'})}{S_{\rm min}(w)} = \frac{N(w^{'})}{N(w)}\quad \]
This conjecture will be tested for the ${\tt a}^n {\tt b}^n$ sequence in Sect. \ref{ss:a_nb_n_sequence},
for ${\tt a} {\tt b} {\tt A}^n {\tt B}^n$ sequence in Sect. \ref{ss:abA_nB_n_sequence} 
and for the ${\tt a} {\tt b} {\tt a}^n {\tt b}^n$ sequence in Sect. \ref{ss:aba_nb_n_sequence}. 

The action S can be directly evaluated as 
\[S_{\rm min} = - {2} \int_{0}^{T} V({\bf r}(t)) d t\]
where ${\bf r}(t)$ represents the periodic solution to the e.o.m. with given energy $E=0$. 
We may factor out the constant term $2 G m^2/R_0^2 = 1$ (due to $G = m = 1;~R_0^2 = 2$)
in front of the integral,
\begin{equation}
S_{\rm min} = {2} \frac{G m^2}{R^2} \int_{0}^{T} V(\phi(t),\theta(t)) d t 
\label{e:S_2V} 
\end{equation}
and evaluate the remaining integral with our solutions. In spite of the explicit
scale ($R$) dependence, this integral is scale invariant, due to the time $t$'s
compensating scale dependence.

\subsection{Action  ``quantization''}
\label{ss:Action_quantization}

\subsubsection{Sequence ${\tt a}^n {\tt b}^n$}
\label{ss:a_nb_n_sequence}

\begin{figure}
\centerline{\includegraphics[width=0.75\columnwidth,,keepaspectratio]{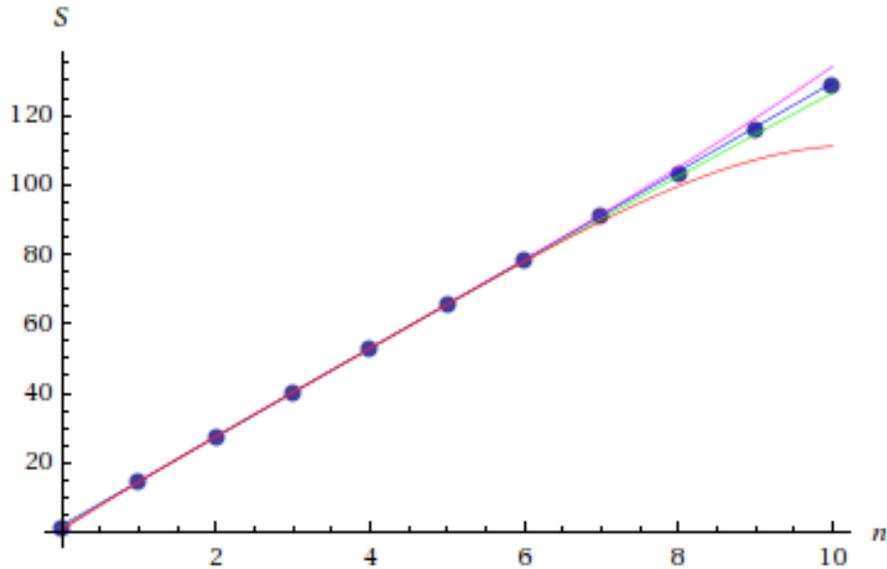}}
\caption{S(n) ``curve'' for the ${\tt a}^n {\tt b}^n$ sequence of solutions with $n$=0,...,10 from 
Table \ref{tab:Action1} together with four different polynomial fits (linear, quadratic, cubic and
quartic (blue, green, magenta and red, respectively) to the $n$=1,...,5 orbits, 
the values at $n=0, 6, \cdots 10$ being predictions. One can see that all four fits predict 
the $n=0$ point correctly, but only the linear and the quadratic fits (blue and green, respectively) 
bracket the higher-$n$ values.} 
\label{fig:Action1b}
\end{figure}

We test the linear action-topology regularity for the ${\tt a}^n {\tt b}^n$ sequence, 
Fig. \ref{fig:Action1b}, with data from Table \ref{tab:Action1}.
At first, we fit the five points ($n \in$ [1,5]) with linear, quadratic, cubic and
quartic polynomials:
The linear fit to orbits with $n \in$ [1,5] is 
${\rm S}_1^{\rm I}(n) = 1.82486 + 12.7547 n$ 
the quadratic one 
\footnote{The cubic fit is
${\rm S}_3^{\rm I}(n) = 0.988914  + 13.6903 n - 0.27658 n^2 + 0.023812 n^3$
and the quartic one
${\rm S}_4^{\rm I}(n) = 0.53508 + 14.5457 n -  0.797349 n^2
+ 0.149877 n^3 -0.0105054 n^4$.}
${\rm S}_2^{\rm I}(n) = 1.38896 +13.1283 n - 0.0622721 n^2$
Then we use these fits to predict: 1) the single lower-lying ($n=0$) orbit's action;
2) the higher-lying ($n \in$ [6,10]) orbits' actions S$(n)$ 
and compare them with the actual values in Table \ref{tab:Predict_Action1}, shown 
in Fig. \ref{fig:Action1b}. One can immediately see that the cubic and the quartic 
fits deviate significantly from the data at higher values of $n \geq 8$, 
Fig. \ref{fig:Action1b}, thus suggesting that all terms with powers higher 
than the second are inappropriate.
\begin{table}
\caption{\label{tab:Action1} The action ${\rm S}$, the period $T$, the difference of action 
$\Delta{\rm S}_n = {\rm S}_n - {\rm S}_{n-1}$, the difference of differences
$\Delta^{2}{\rm S}_n = \Delta{\rm S}_n - \Delta{\rm S}_{n-1}$ 
and the free group element (f.g.e.) of three-body orbits in
the ${\tt a}^n {\tt b}^n$ sequence. The entry denoted by 0 is the Calogero-Schubart colliding orbit
whose period and action were calculated analytically in 
\ref{s:Calogero}.}
\footnotesize
\begin{tabular}{@{}lllll}
\br
${\rm T}$  & ${\rm S}$ & $\Delta{\rm S}_n$ & $\Delta^{2}{\rm S}_n$ & {\rm f.g.e.} \\
\mr 
0.88889 & 1.09861 & - & - & ${\tt a^0 b^0}$ \\ 
\mr
2.4254 & 14.4191 & 12.2256 & - & ${\tt a^1 b^1}$ \\ 
2.82025 & 27.4681 & 13.0453 & 0.8197 & ${\tt a^2 b^2}$ \\ 
2.92057 & 40.231 & 12.7238 & -0.3215 & ${\tt a^3 b^3}$ \\ 
2.96001 & 52.9039 & 12.6713 & -0.0525 & ${\tt a^4 b^4}$ \\ 
2.97947 & 65.5371 & 12.6356 & -0.0357 & ${\tt a^5 b^5}$ \\ 
2.99047 & 78.1493 & 12.547 & -0.0886 & ${\tt a^6 b^6}$ \\  
2.9973 & 90.7493 & 12.7089 & 0.1619 & ${\tt a^7 b^7}$ \\ 
3.00183 & 103.341 & 12.5883 & -0.1206 & ${\tt a}^8 {\tt b}^8$ \\ 
3.00498 & 115.928 & 12.587 & -0.0013 & ${\tt a^9 b^9}$ \\ 
3.00727 & 128.51 & 12.582 & -0.005 & ${\tt a^{10} b^{10}}$ \\ 
\mr
\end{tabular}
\end{table}

Thence we see that:
1) all four fits agree with/correctly predict the order of magnitude 
of the $n=0$ (the Calogero-Schubart orbit's) action, where the 
calculated value ${\rm S}(0) = 1.09861$, see 
\ref{s:Calogero}, 
satisfies the inequality ${\rm S}_2(0) \leq {\rm S}(0) \leq {\rm S}_3(0)$;
2) the largest deviations of $\Delta{\rm S}_n$ and $\Delta^{2}{\rm S}_n$ from 
their mean values are for $n = 0,1,2$, the calculation of which involves $S(0)$.; 
3) the actual values for $n \in$ [6,10] generally fall between the linear 
and the quadratic fits: 
${\rm S}_1(n) \leq {\rm S}(n) \leq {\rm S}_2(n)$, see Table \ref{tab:Action1}, 
and Fig. \ref{fig:Action1b}.

Note that the constraint 
$12.7547  \leq c_1 \leq 13.1283$ is  
in (rough) agreement with the $\Delta S^{\rm I}$ in Table \ref{tab:Action1}, 
and that the fitted value of $c_2 \simeq - 0.0622721$ generally 
agrees with the sign and the order of magnitude of $\Delta^2 S$. As $n$ 
increases, the values of $\Delta{\rm S}_n$ and $\Delta^{2}{\rm S}_n$ approach 
constants. All of this suggests that the 
value of the action $S(n)$ asymptotically approaches
a linear function of $n$, as $n \to \infty$.

The linear fit to all 11 orbits in sequence I, 
$S_{1}^{\rm I}(n) = 1.78211 + 12.7004 n$
and the complete quadratic one, 
$S_{2}^{\rm I}(n) = 1.3729 + 12.9732 n - 0.0272804 n^2$
are in slightly better agreement with the data than the ones without the first term.
\begin{table}
\caption{\label{tab:Predict_Action1} Predictions (linear ${\rm S}_1$, quadratic ${\rm S}_2$ , cubic ${\rm S}_3$), 
of the action S of three-body orbits in the ${\tt a}^n {\tt b}^n$ sequence. The first orbit 
($n=0$) is the Calogero-Schubart colliding one.}
\footnotesize
\begin{tabular}{@{}llllll}
\br
${\rm T}$ & S & ${\rm S}_1$ & ${\rm S}_2$  & ${\rm S}_3$ & {\rm f.g.e.} \\
\mr
0.88889 & 1.09861 & 1.82486 & 1.38896 & 0.988914 & ${\tt a}^0 {\tt b}^0$ \\ 
2.99047 & 78.1493 & 78.3531 & 77.9172 & 78.3172 & ${\tt a}^6 {\tt b}^6$ \\ 
2.9973 & 90.7493 & 91.1078 & 90.236 & 91.4361 & ${\tt a}^7 {\tt b}^7$ \\ 
3.00183 & 103.341 & 103.862 & 102.43 & 105.002 & ${\tt a}^8 {\tt b}^8$ \\ 
3.00498 & 115.928 & 116.617 & 114.5 & 119.158 & ${\tt a}^9 {\tt b}^9$ \\ 
3.00727 & 128.51 & 129.372 & 126.445 & 134.046 & ${\tt a}^{10} {\tt b}^{10}$ \\ 
\mr
\end{tabular}

\end{table}

\subsubsection{Sequence ${\tt a} {\tt b} {\tt A}^n {\tt B}^n$}
\label{ss:abA_nB_n_sequence}

\begin{figure}
\centerline{\includegraphics[width=0.75\columnwidth,,keepaspectratio]{
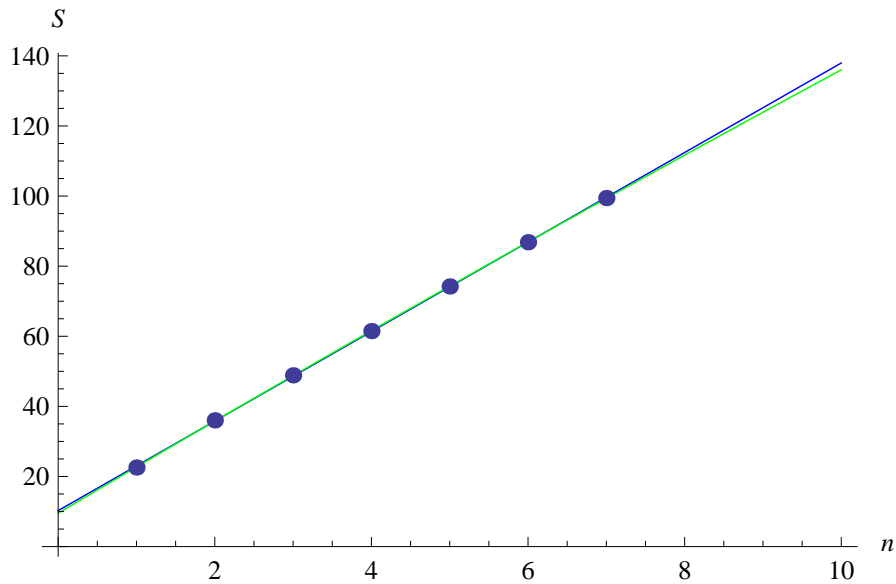}}
\caption{The action S as a function of $n$ 
for the ${\tt a} {\tt b}{\tt A}^n {\tt B}^n$ sequence of solutions with $n$=0,...,7 from 
Table \ref{tab:Action2} together with two lowest-order polynomial fits
(the same color code as in Fig. \ref{fig:Action1b}).} 
\label{fig:Action2}
\end{figure}

We repeat this procedure for the ${\tt a} {\tt b} {\tt A}^n {\tt B}^n$ sequence
and find similar results in Table \ref{tab:Action2} and Fig. \ref{fig:Action2}.
\begin{table}
\caption{\label{tab:Action2} The action ${\rm S}$, the period $T$, the difference of action 
$\Delta{\rm S}_n = {\rm S}_n - {\rm S}_{n-1}$, the difference of differences
$\Delta^{2}{\rm S}_n = \Delta{\rm S}_n - \Delta{\rm S}_{n-1}$ 
and the free group element (f.g.e.) of three-body orbits in
the ${\tt a}^n {\tt b}^n$ sequence (II). The entry denoted by (${\tt ab A^0 B^0}$) is  
also the first orbit from sequence I.}
\footnotesize
\begin{tabular}{@{}lllll}
\br
${\rm T}$  & ${\rm S}$ & $\Delta{\rm S}_n$ & $\Delta^{2}{\rm S}_n$ & {\rm f.g.e.} \\
\mr
2.4254 & 14.4191 & - & - & ${\tt ab A^0 B^0}$ \\ 
\hline
5.02974 & 22.5645 & 8.1454 & - & ${\tt ab A^1 B^1}$ \\ 
5.97849 & 36.0148 & 13.4503 & 5.3049 & ${\tt a b A^2 B^2}$ \\ 
6.14445 & 48.8613 & 12.8465 & -0.6038 & ${\tt a b  A^3 B^3}$  \\ 
6.18929 & 61.4806 & 12.6193 & -0.2272 & ${\tt a b A^4B^4}$ \\
6.22509 & 74.2112 & 12.7306 & 0.1113 & ${\tt a b A^5 B^5}$ \\
6.23852 & 86.8308 & 12.6196 & -0.111 & ${\tt a b A^6 B^6}$ \\
6.2463 & 99.433 & 12.6022 & -0.0174 & ${\tt a b A^7 B^7}$ \\
\mr
\end{tabular}
\end{table}

The linear fit, excluding the $n=0$ point, is 
$S_1^{\rm II}(n) = 10.2584 + 12.771 n$
the quadratic one 
$S_2^{\rm II}(n) =  9.52234 + 13.2617 n - 0.0613381 n^2$. 
Including the $n=0$ point, the fits become
$S_1^{\rm II}(n) = 11.992 + 12.4243 n$
and
$S_2^{\rm II}(n) =  12.9909 + 11.4254 n + 0.142693 n^2$.
Note that the magnitudes of $c_1, c_2$ in these fits generally agree 
with those in the sequence ${\tt a}^n {\tt b}^n$,
with $c_0$ being an exception, for obvious reasons.

\subsubsection{Sequence ${\tt a} {\tt b} {\tt a}^n {\tt b}^n$}
\label{ss:aba_nb_n_sequence}

\begin{figure}
\centering\includegraphics[width=0.75\columnwidth,,keepaspectratio]{
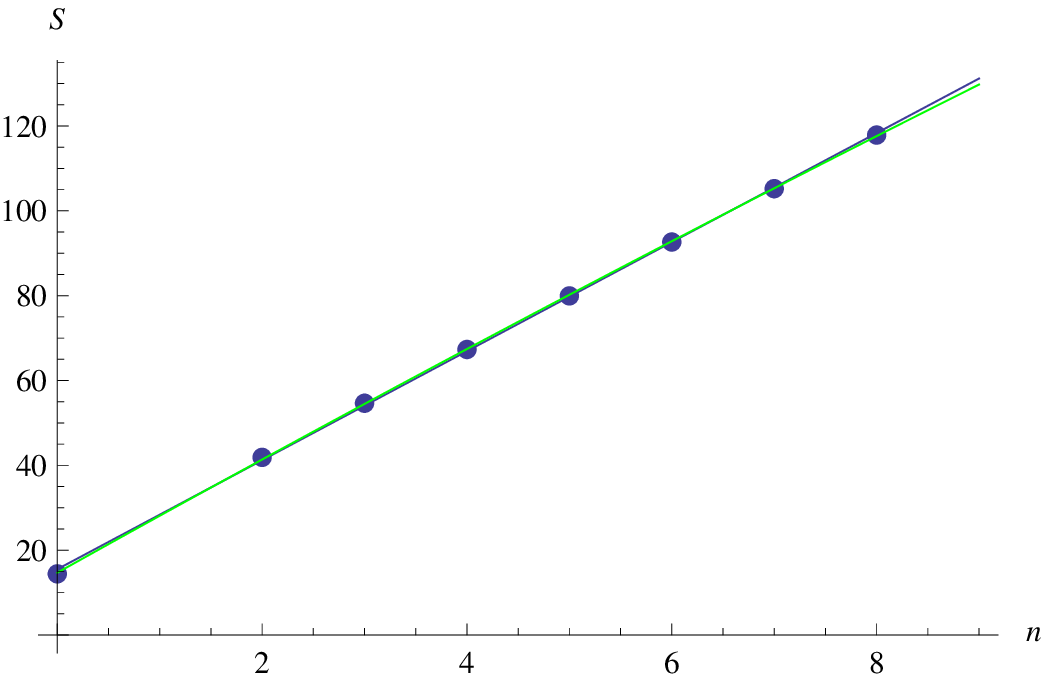}
\caption{The action S as a function of $n$ for the ${\tt a} {\tt b}{\tt a}^n {\tt b}^n$ sequence of  
solutions with $n$=0,2,...,8 from Table \ref{tab:Action3} together with two lowest-order 
polynomial fits (the same color code as in Fig. \ref{fig:Action1b}).} 
\label{fig:Action3}
\end{figure}

We repeat this procedure for the ${\tt a} {\tt b} {\tt a}^n {\tt b}^n$ sequence
and find similar results in Table \ref{tab:Action3} and Fig. \ref{fig:Action3}.
\begin{table}
\caption{\label{tab:Action3} The action ${\rm S}$, the period $T$, the difference of action 
$\Delta{\rm S}_n = {\rm S}_n - {\rm S}_{n-1}$, the difference of differences
$\Delta^{2}{\rm S}_n = \Delta{\rm S}_n - \Delta{\rm S}_{n-1}$ and the free group element 
(f.g.e.) of three-body orbits in
the ${\tt a b a}^n {\tt b}^n$ sequence. The entry denoted by 0 is the Calogero-Schubart colliding orbit
whose period and action were calculated analytically in Appendix \ref{s:Calogero}.}
\footnotesize
\begin{tabular}{@{}lllll}
\br
${\rm T}$  & ${\rm S}$ & $\Delta{\rm S}_n$ & $\Delta^{2}{\rm S}_n$ 
& {\rm f.g.e.} \\
\br 
2.4254 & 14.4191 & - & - & ${\tt a^0 b^0 ab}$ \\ 
\br
5.23969 & 41.8824 & 13.7316 & - & ${\tt a^2 b^2 a b}$ \\ 
5.33796 & 54.6432 & 12.7608 & -0.97085 & ${\tt a^3 b^3 a b}$ \\ 
5.37678 & 67.3197 & 12.6765 & -0.0843 & ${\tt a^4 b^4 a b}$ \\ 
5.39600 & 79.9482 & 12.6285 & -0.048 & ${\tt a^5 b^5 a b}$ \\ 
5.40845 & 92.6465 & 12.6983 & 0.0698 & ${\tt a^6 b^6 a b}$ \\ 
5.41368 & 105.229 & 12.5825 & -0.1158 & ${\tt a^7 b^7 a b}$ \\ 
5.41817 & 117.864 & 12.6350 & 0.0525 & ${\tt a^8 b^8 a b}$ \\ 
\br 
\end{tabular}
\end{table}
The linear fit, excluding the $n=0$ point, is 
$S_1^{\rm III}(n) = 16.6399 + 12.6587 n$
the quadratic one 
$S_1^{\rm III}(n) =  16.4 + 12.7729 n - 0.0114214 n^2$. 
Including the $n=0$ point, the fits become
$S_1^{\rm III}(n) = 15.51 + 12.8535 n$
and
$S_1^{\rm III}(n) =  14.691 + 13.5012 n - 0.0793937  n^2$. 
Once again the same remarks apply: the values 
of $c_1, c_2$ in these fits are generally close to those in the previous two sequences. 
The manifest question is: why?

\section{Analytical arguments}
\label{ss:complex}

Calculus with one complex variable has been used to evaluate the action of the 
two-body problem, see Refs. \cite{Landau,Pars:1964}. Application of 
few complex variables methods to the planar few-body problem(s) in celestial 
mechanics can be traced back (at least) to Siegel and Moser \cite{Siegel}. That
means two complex variables for the planar three-body problem. 
In the case of the strong potential, conservation of the hyper-radius, together with the 
zero angular momentum condition, eliminate one of two complex variables, see 
\ref{s:analytic_evid}.
\footnote{Analyticity of solutions to the three-body equations of motion in the Newtonian potential, 
has been proven by Sundman, Ref. \cite{Sundman:1907,Henkel:2001}, subject to the condition 
that there are no three-body collisions, albeit allowing for two-body collisions. 
Analogous proofs do not exist in the case of the strong potential, to our knowledge, but we shall 
nevertheless proceed, subject to the assumption that the periodic solutions to the strong potential 
e.o.m. are also analytic, at least within some region of the complex plane, that is, perhaps, 
determined by collisions. As we deal only with collisionless orbits here, we believe to be 
within the assumed region of analyticity.}

The action of a periodic orbit can be written as a line integral over a closed path on a shape sphere
parametrized by two angles $(\phi(t),\theta(t))$ in Eq. (\ref{e:S_2V}).
Such an integral over the shape-sphere variables can be recast as a Cauchy closed-contour
integral over a single complex variable, see pp. 22-30 in Ref. \cite{Dmitrasinovic:2014}, or 
\ref{s:analytic_evid}, with three poles, because the sphere
can be stereographically projected into a (real) plane, and the (real) plane can be replaced by a
single complex variable \cite{Dmitrasinovic:2014}. Thus we have a closed line integral in a complex 
plane with three poles (one of which is at infinity) - its value is determined by Cauchy's residue
theorem and by the topology of the integration path as follows: 
\begin{eqnarray}
S_{\rm min}(n) &=& S_{\rm min}(0) + 2 i \pi \sum_{\rm poles} {\rm Res} f(n)
\label{e:Res} \
\end{eqnarray}
Then, wee see that the value of the integral is naturally proportional to the number $n$ of times 
the path goes around either of two poles, which is the basic hallmark of our numerical results. 
\begin{eqnarray}
S_{\rm min}(n) 
&\simeq& S_{\rm min}(0) + 2 n i \pi \sum_{\rm poles} {\rm Res}f(1) + {\cal O}(n^2),
\label{e:Res2} \
\end{eqnarray}
where the ${\cal O}(n^2)$ term(s) arise (only) if the difference 
$\Delta {\rm Res}f(n) =  {\rm Res}f(n) - {\rm Res}f(n-1) \neq 0$ between the residues at the $n$-th
and the $(n-1)$st turn does not equal zero.

Note, however, that: (a) there is an additive constant $S_{\rm min}(0) = c_0 \simeq 1$ in the fits
to the sequence (I), 
above - what is its source in terms of complex integration? 
(b) the non-vanishing value of the quadratic term $c_2 \neq 0$ in the fit(s) indicates that 
the value of the residue ${\rm Res}f(n)$ changes with $n$, yielding the ${\cal O}(n^2)$ term - 
but, why should that happen?

(a) The contribution $S_{\rm min}(0) = c_0 \simeq 1$ in sequence (I) simply cannot originate from 
a pole, as its contribution would have to be multiplied by $n$, and consequently would vanish at $n=0$.
This is not necessarily true for sequences (II) and (III), as their free-group elements, 
${\tt a} {\tt b} {\tt A}^n {\tt B}^n$ and ${\tt a} {\tt b} {\tt a}^n {\tt n}^n$, respectively, 
indicate that their $n=0$ terms still involve some contribution from the two poles in the 
complex plane (without the point at infinity).

In the case of sequence (I), $c_0$ can be due to the pole at infinity, which is circumscribed 
only once, or from crossing a branch cut. 
A branch cut would also explain the different changes of the residue with the changing $n$, but 
that cut would have to be logarithmic in order to change the residues at infinitely many values of 
$n \in [0, \infty]$. 
Now, no branching points are manifest in Eq. (\ref{e:S_2V}), but two have been explicitly found, 
associated with the two-body collison poles, and the logarithmic nature of the branch cut shown, 
by a calculation of the action of one particular solution, see 
\ref{s:Calogero}.

In 
\ref{s:Calogero} we have evaluated the action of the collisional zero-angular momentum 
Calogero-Schubart orbit, 
Refs. \cite{Calogero:1969xj,Khandekar:1972,Calogero2001,Hakobyan:2009ac,Hakobyan:2010ia},
and show that it provides: 1) a logarithmic branch cut;
2) the correct size 
of the constant term $S_{\rm min}(0) = c_0$ in the 
power series, that is related to the discontinuity of the integral at the branch cut.

(b) The size of the change of the residue $\Delta {\rm Res}f(n) =  {\rm Res}f(n) - {\rm Res}f(n-1)$
due to the ``crossing of the cut'' can not be estimated with our present knowledge, but 
if it 
is small compared with the actual value of the residue, 
$\Delta {\rm Res}f(n) <<  {\rm Res}f(n)$, then it would explain both 
(1) the approximately equal slopes $c_1 \simeq {\rm Res}f(n) $ of the $S(n)$ graphs, 
for different sequences of orbits; and 
(2) the branch cut implies a non-vanishing quadratic term $c_2 \simeq \Delta {\rm Res}f(n) \neq 0$
that varies from one value of $n$ to another.
The above arguments suggest that the analyticity of solutions 
could be sufficient to explain the remarkable, numerically observed behaviour of the action.


\section{Summary and Conclusions}
\label{s:summary}

We have searched for periodic planar three-body orbits in the strong three-body problem, and 
found 24 orbits, 22 of which are new. These solutions neatly fill in the lowest entries 
in Montgomery's topological classification of three-body orbits  
and contain several solutions with topologies that do not exist in the 
Newtonian potential. These 24 orbits fall into three algebraically well-defined sequences.

Two properties of the discovered orbits shine through immediately:
1) periodic orbits exist only in (several) small regions of the allowed phase space;
2) the orbits discovered thus far form three sequences with respect to the algebraic description
of their topologies; 3) the orbits' periods $T$ and initial conditions 
converge to, in this case three (thus far) accumulation points, which makes them difficult 
to disentangle beyond some fairly small number (about 10 when working to a precision of 
16 decimal places) of orbits in each sequence. We are not aware of another system of 
ordinary differential equations that has a similar structure of solutions. 

Thus we have shown that the problem of finding zero-angular-momentum periodic three-body orbits 
in the strong potential is amenable to a numerical study, albeit with some perhaps unexpectedly 
severe limitations imposed by the structure of the solutions themselves. 
The task of finding an efficient method of disentangling further sequences of orbits 
and their accumulation points remains as a challenge for future work. 

The new orbits' periods depend on the specific sequence of orbits, but always remain bounded from 
above, approaching different asymptotic limits, thus explicitly contradicting the conjectured 
linear relationship between the scale-invariant period and the index $n$ of the sequence.
Furthermore, in each sequence there is a remarkable almost linear relation between the 
initial angle $\phi_n$ and the period $T_n$, as $n \to \infty$.
The orbits' configuration-space trajectories fall within compact, clearly separated 
regions in the plane and have both an inner and an outer envelope. 
All of this put together might indicate a certain structure to the set of all periodic 
solutions, if not an outright solvability of the problem.

We have calculated the values of the (scale invariant) action, 
which equals twice the time integral over the period of the kinetic energy,  
for these periodic orbits and found that they are ``topologically quantized'', 
i.e., approximately linearly proportional to the index $n$ of the sequence (or 
equivalently to the number of ``letters'' in the ``words'' describing their topologies), 
with a small quadratic correction that diminishes with increasing $n \to \infty$. 
The existence of both of these features of the action can be explained in terms of 
Cauchy's theorem for a closed contour integral. 

Our results ought to have interesting consequences
for the two physical systems that contain the strong three-body potential: 
(1) the post-Newtonian gravity; and (2) semiclassical calculation of the Efimov effect. 
Unfortunately, our present results are insufficient to draw conclusions about these 
two physical systems, as yet. This calls for new and different methods to be invented
and applied to this system.

\ack{}
The work of V.D. and M.\v S was financed by the Serbian Ministry of 
Science and Technological Development under grant numbers OI 171037 and III 41011.
The work of L.V.P. was conducted during the summer vacations of 2013 and 2014,
and was 
accepted in lieu of the Praktikum in the ``Scientific Computing'' 
module of the third year of 
undergraduate curriculum at the Faculty of Physics of the University of Vienna.
All numerical work was done on the Zefram cluster, Laboratory for gaseous electronics,
Center for non-equilibrium processes, at the Institute of Physics, Belgrade.

\appendix

\section{Numerical accuracy of initial conditions for periodic orbits in the ``strong'' $-1/r^2$ potential}
\label{s:Numerical}

\begin{figure}
\hskip48pt
\includegraphics[width=0.85\columnwidth,,keepaspectratio]{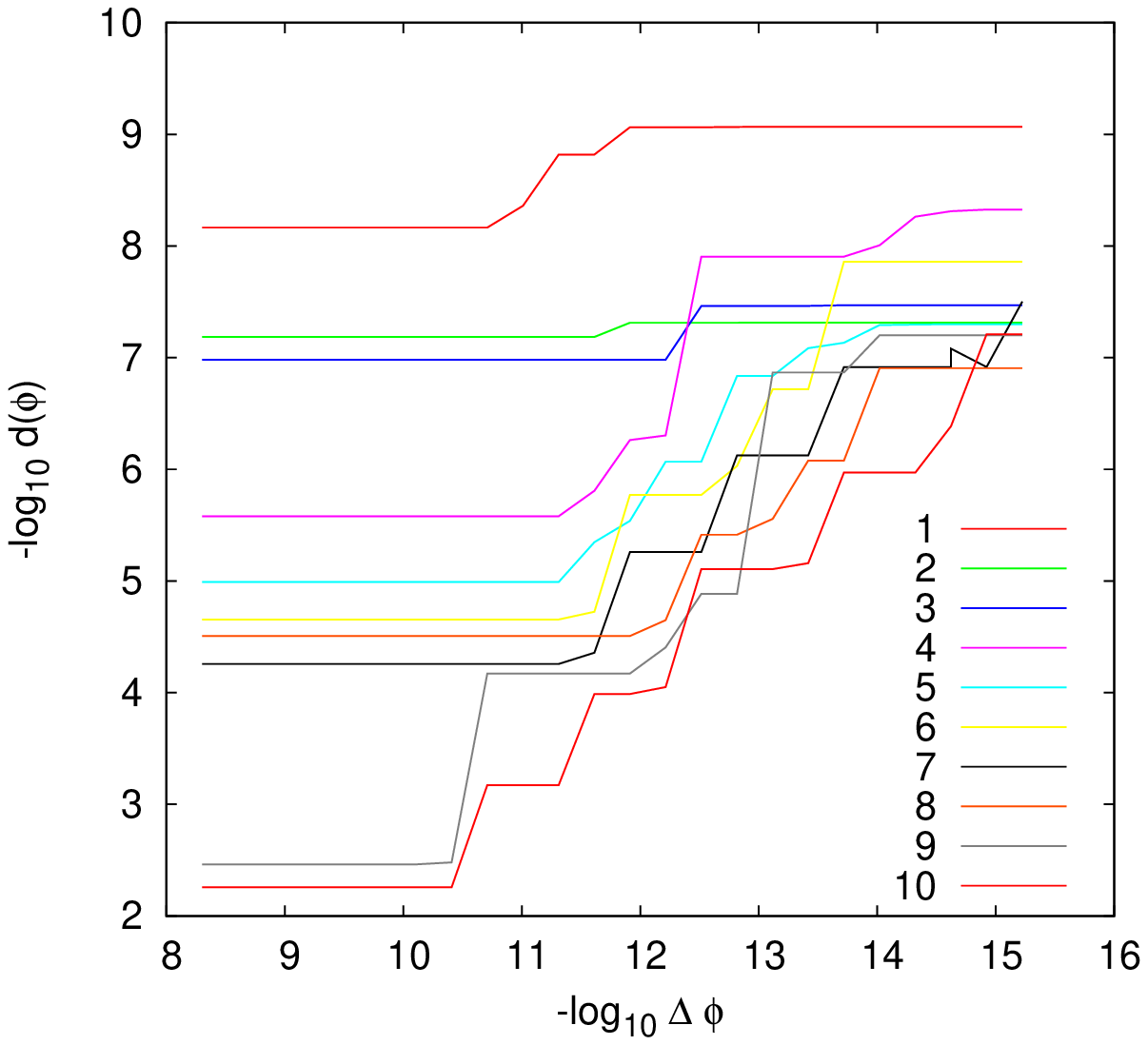}
\vskip40pt
\caption{The negative logarithm of the return proximity function $−\log(d)$ as a function of 
the negative logarithm of the deviation (``error'') $\Delta \phi_n = |\phi_n - \phi_n(0)|$ of the 
initial angle $\phi_n$ from the exact value $\phi_n(0)$ for the first ten periodic orbits, $n=1, \ldots ,10$
in sequence I. Here one can graphically see the importance of the higher-lying decimal points for the 
value of the return proximity function.}
\label{fig:descentplot1}
\end{figure}

\begin{figure}
\centering
\hskip4pt
\includegraphics[width=0.45\columnwidth,,keepaspectratio]{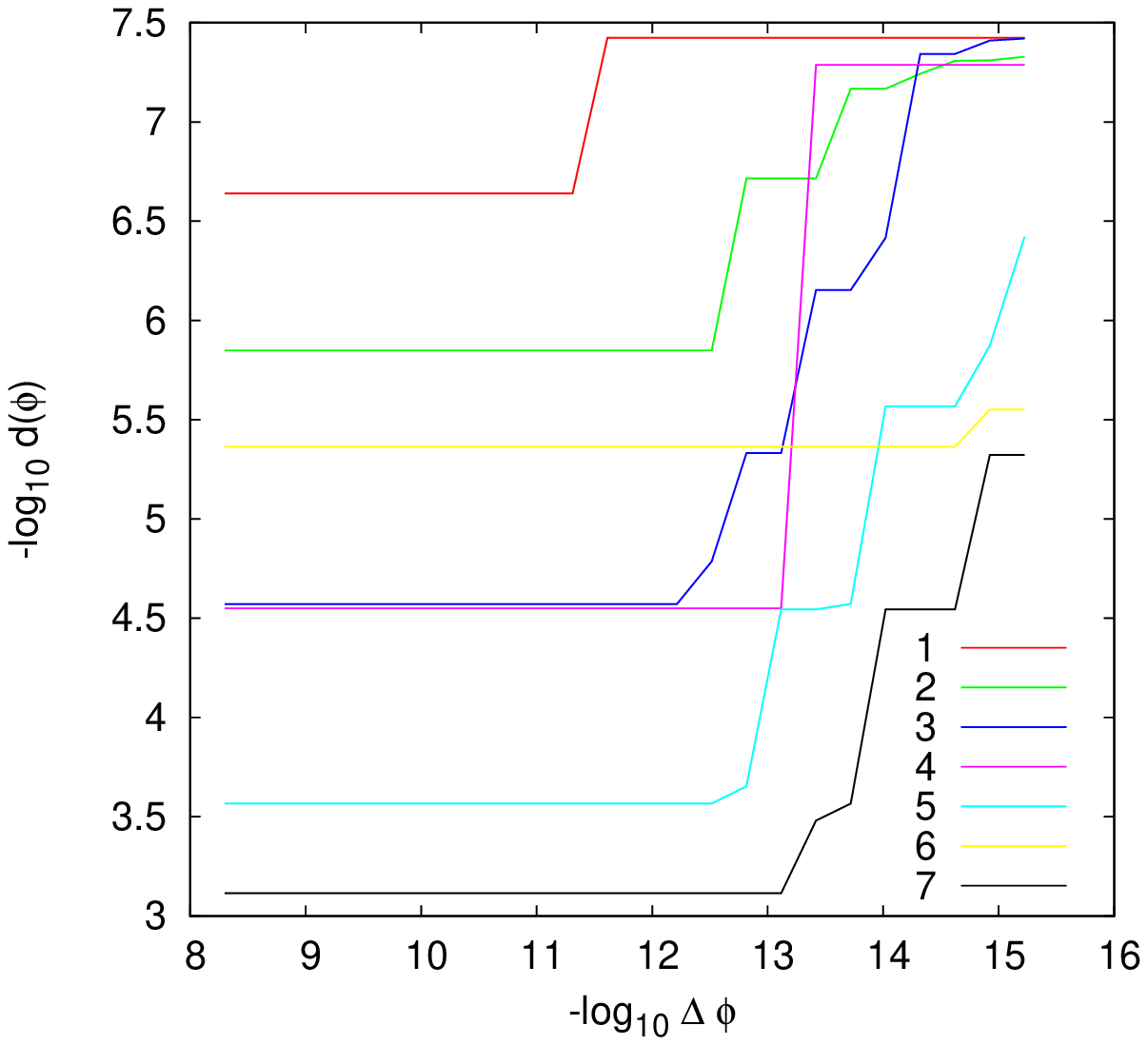},
\includegraphics[width=0.45\columnwidth,,keepaspectratio]{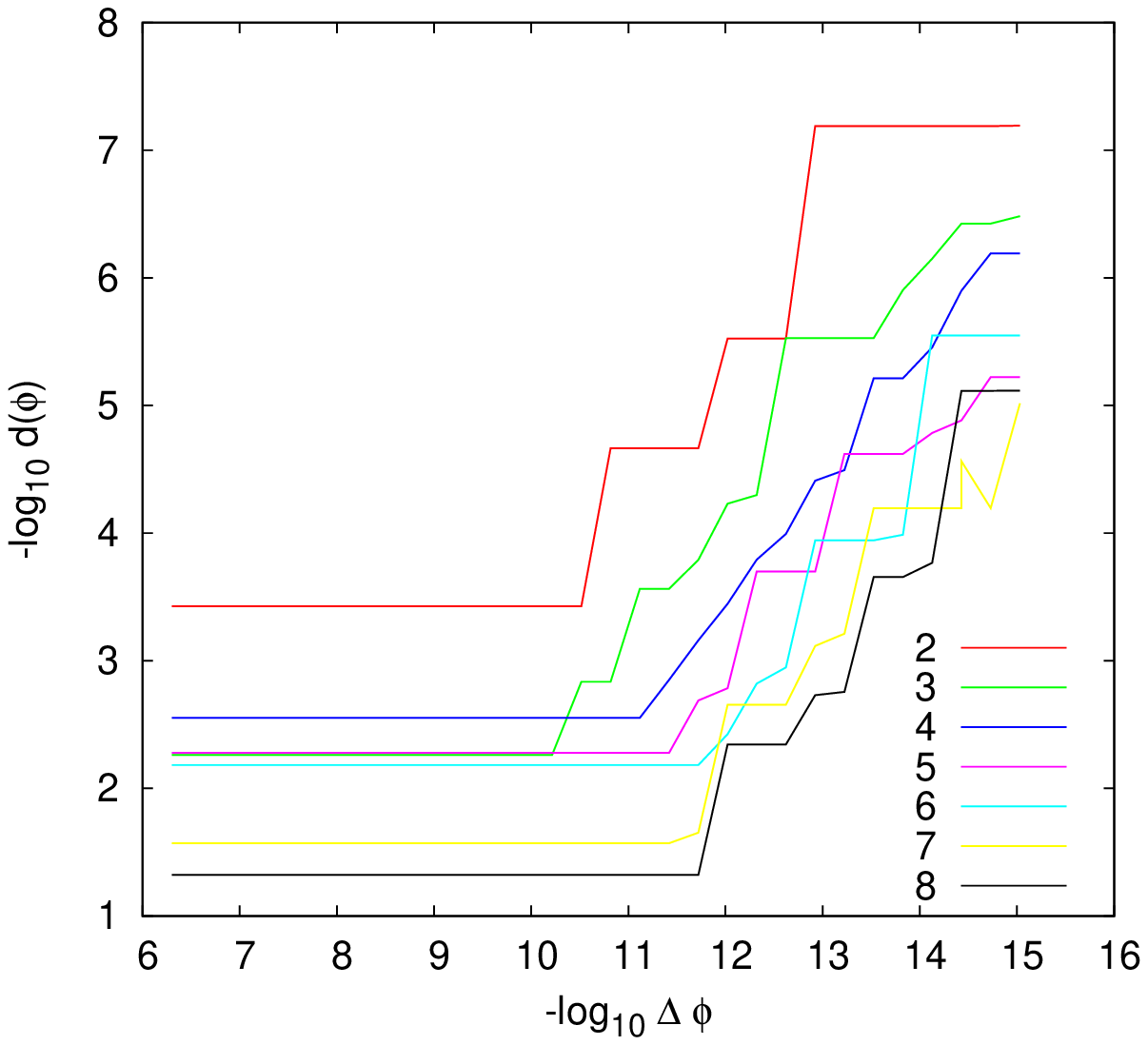}
\vskip24pt
\caption{Same as in Fig. \ref{fig:descentplot1}, only for sequences II and III.}
\label{fig:descentplot23}
\end{figure}

The negative logarithm of the deviation $\Delta \phi = |\phi - \phi_0|$ of the approximate
initial angle $\phi$ from the exact value $\phi_0$ of a periodic orbit 
represents the highest decimal place of the ``error'' committed.
In Figs. (\ref{fig:descentplot1}),(\ref{fig:descentplot23}) 
we show the dependence of the (negative logarithm of the) return proximity function
as a function of the negative logarithm of the ``error'' $\Delta \phi$ for periodic orbits 
in the three sequences, respectively, during the process of minimization of the return proximity
function using the gradient descent procedure. 
There one can see that generally speaking one needs an accuracy of about 
14 decimal places in order to achieve uniform return proximity function values within
one sequence.
Note that the ``threshold'' (minimal value of $\Delta \phi_n = |\phi_n - \phi_n(0)|$
at which the $-\log(d)$ starts growing) moves from 10 to 13, depending on the orbit ($n$) and 
the sequence. Moreover, note that some orbits, such as $n=1$ in sequence I, have fairly
large values of $-\log(d)$ even at fairly low values of $-\log\Delta \phi_n$.

\section{Action of the periodic three-body orbit in the one-dimensional ``strong'' $-1/r^2$ potential}
\label{s:Calogero}

In the one-dimensional (1D) ``strong'' $-g/r^2$ potential Calogero-Moser model, Refs. 
\cite{Calogero:1969xj,Calogero2001,Moser1975}, the periodic three-body problem has only 
one variable 
- the meridional angle (``longitude'') $\phi$ of a point on the equator of the shape-sphere, 
see Refs. \cite{Calogero:1969xj,Khandekar:1972,Calogero2001,Hakobyan:2009ac,Hakobyan:2010ia}. 
This orbit corresponds to the collinear v. Schubart orbit in Newtonian gravity.


\subsection{Solving the equation of motion}
\label{ss:Energy conservation}
 
Energy conservation $E =0$ yields as the equation of motion
\[E = 
\frac{m}{8} R^2 {\dot \phi}^2
- \frac{9}{2} \frac{G m^2}{R^2} \frac{1}{\sin^2\left(\frac{3 \phi}{2}\right)} = 0 \]
which is readily solved as 
\[ {\dot \phi}(t) = \frac{d \phi}{d t} = \pm \frac{6 \sqrt{2 g}}{R^2 \sqrt{1-\cos (3 \phi )}} \]
where $g = G m$.
Integrate this equation over $d \phi$ to find 
\[t - t_0 = \frac{R^2}{6 \sqrt{2 g}} \int \sqrt{1-\cos (3 \phi )} \, d\phi \]
\[= -\frac{R^2}{9\sqrt{2 g}}  \sqrt{1-\cos (3 \phi )} \cot \left(\frac{3 \phi}{2}\right) \]
which, after the simplification
\[\sqrt{1 - \cos(3 \phi)} \cot \left(\frac{3 \phi }{2}\right) = \sqrt{2} \cos\left(\frac{3 \phi }{2}\right)\]
leads to
\[t - t_0 = \mp \frac{R^2}{9\sqrt{g}} \cos\left(\frac{3 \phi }{2}\right) \]
Note the ``dimensionless time'' variable $t^{'} = \frac{9 \sqrt{g}}{R^2}t$.
This is identical to 
\begin{equation}
\frac{3 \phi}{2} = {\rm Arccos} \left( \mp \frac{9\sqrt{g}}{R^2} \left(t - t_0\right)\right) .
\label{e:Ansatz}
\end{equation} 
Setting $t_0 = 0$ we have finally 
\[t^{'} =  \frac{9\sqrt{g}}{R^2} t  =  \mp \cos \left(\frac{3 \phi}{2} \right) .\] 
as the solution. 

\subsubsection{Solution with $g = G m = 1, R^2 = 2$}

In the $g = G m = 1, R^2 = 2$ limit, the time $t^{'} =  \frac{9\sqrt{g}}{R^2} t = \frac{9}{2} t$ 
and the e.o.m. turns into
\[{\ddot \phi}(t) = \frac{486 t}{\left(4 - 81 t^2\right)^{3/2}} \]
The solution to this equation is
\[\phi(t) = \frac{1}{3} \left(3 c_2 t + 2 {\rm Arcsin} \left(\frac{9t}{2}\right)- \pi \right)\]
where $c_2$ is an integration constant; setting $c_2 = 0$ we find:
\begin{equation}
\phi(t) + \frac{1}{3}\pi = \frac{2}{3} {\rm Arcsin} \left(\frac{9t}{2}\right)~~ 
\label{e:**} \
\end{equation}
which is equivalent 
to the Ansatz Eq. (\ref{e:Ansatz}).
Note that the equation (\ref{e:**}) does not have real solutions for time $t$ larger than $t > 2/9$,
as the argument of ${\rm Arcsin} = \sin^{-1}$ exceeds unity. This means that the $\phi$ turns 
imaginary at the $t = 2/9$ point in time, i.e., $t = 2/9$ is a branch point of the solution.

We demand that $\phi$ be real, which implies an upper and a lower limit on the (real) time 
variable $t$:
\[-1 \leq \frac{9t}{2} \leq 1 .\]
For other times, the solution to  the Ansatz Eq. (\ref{e:Ansatz}) becomes imaginary/complex.
These two ``boundaries'' on time $t$ imply the following limits on the value of $\phi$:
\[\phi_{\rm max} = \phi(2/9) = -\frac{2}{3} {\rm Arccos}\left(1\right) \equiv 0 ~({\rm mod} 2\pi) \]
\[\phi_{\rm min} = \phi(-2/9) = -\frac{2}{3} {\rm Arccos}\left(-1\right)= - \frac{2}{3} \pi \]
Thus, one cycle consists of an oscillation of $\phi$ from $\phi_{\rm min} = \phi (-2/9) = 
- \frac{2}{3} \pi $ 
to $\phi_{\rm max} = \phi (2/9) = 0$ and back. Therefore, the half-period $T/2$ is
\[T/2 =  2/9 - (-2/9) = 4/9 \]

\subsection{Action of the Schubart-Calogero-Moser solution}
\label{ss:Action_SCM}

The action S can be evaluated from either the time integral over one period of twice the kinetic 
energy, $2 T  = \left(\frac{R^2}{2} p_{\phi}\right)^2$, or the negative of twice the potential 
energy $V = \frac{- 9 g}{2 R^2 \sin^2 \frac{3}{2} \phi}$, where ($g = G m^2 = 1$), see 
Refs. \cite{Khandekar:1972,Hakobyan:2009ac,Hakobyan:2010ia}. 
\[S_{\rm min}(T) = 2 \int_{0}^{T} T( {\dot {\bf r}}(t)) d t = - 2 \int_{0}^{T} V({\bf r}(t)) d t\]

The action evaluated over one period $T$ with the solution Eq. (\ref{e:Ansatz}) equals formally 
\[S_{\rm min}(T) = - \int_{0}^{8/9} \frac{18}{4-81 t^2} \, dt  ~,\]
but manifestly the integrand has poles at $4 - 81 t^2 = 0$, i.e., at the values $t=\pm \frac{2}{9}$ 
of the integration variable $t$, and the integral is not defined for $t > \frac{2}{9}$ due to the 
$t$ turning imaginary (Eq. (\ref{e:Ansatz}) implies a logarithmic branch cut in the complex-$t$ 
plane).

Therefore, we must either 1) make an analytic continuation of the solution for time values $t > \frac{2}{9}$;
or 2) reformulate the integral within the confines of the real solution $\phi(t)$.

The action of a periodic orbit $S_{\rm min}(T)$ must be:
1) independent of the starting (and ending point), i.e.,
\[S_{\rm min}(T) = 2 \int_{-T/2}^{T/2} T( {\dot {\bf r}}(t)) d t 
= - 2 \int_{-T/2}^{T/2} V({\bf r}(t)) d t \]
2) equal to twice the action over one half-period $T/2$
\[S_{\rm min}(T) = 2 S_{\rm min}(T/2) = - 4 \int_{0}^{T/2} V({\bf r}(t)) d t\]
due to the symmetry under time reversal of the integrand.
The integral 
\[S_{\rm min}(T) =  - \int_{-4/9}^{4/9} \frac{18}{4-81 t^2} \, dt  
=  - 2 \int_{0}^{4/9} \frac{18}{4-81 t^2} \, dt\]
is singular: it has a simple pole within the 
integration range. 
When we change the integration variable $t^{'} = z = \frac{9}{2} t$ we find
\[S_{\rm min}(T) = - 2 \int_{0}^{2} \frac{\, dz}{1 - z^2} 
= - 2 \int_{0}^{2} \frac{\, dz}{(1 - z)(1 + z)}  \]
We see that the pole at $z=1$ sits in the middle of the integration
range, but the integral can be evaluated in (at least) two different ways:
a) by elementary real integration; 
b) by complex integration.
Both ways will be useful so as to understand the result.

a) The integral can be evaluated as 
\begin{eqnarray}
S_{\rm min}(T) &=& 
\left(\log (-z-1)- \log (z-1) \right) \Big|_{z=0}^{z=2} 
\nonumber \\
&=& \log(-3) = \log(3) \pm i \pi  
\label{e:S} \ 
\end{eqnarray}
Here, in the last step one may take either sign in evaluating $\log(-1) = \pm i \pi$;  
this choice of sign fixes the phase convention. The said convention determines 
the location of branch cut(s) in the complex plane, which is not immediately obvious. 

b) The imaginary part of the action indicates that the complex integration path 
has crossed a branch cut: the poles at $z=\pm 1$ in Eq. (\ref{e:S}) are also logarithmic 
branch points, each with its own branch cut.
Logarithmic branch cuts are usually fixed so as to extend from the branch point 
(the zero of the argument of the logarithm) to infinity, but that is not compulsory - 
the cuts may be chosen in other ways, as well.
The real part of the action is invariant under the change of the location of branch cuts; 
by changing the branch cuts one can only change 
the sign of the imaginary part, which, in turn, is equivalent to a change of integration contour.

The numerical value of the real part of the minimized action is  
\begin{equation}
S_{\rm min}(T) = 1.09861 
\end{equation}
which is in good agreement with the rest of the graph in Fig. \ref{fig:Action1b}.

\section{Analytic properties of the action and complex variables}
\label{s:analytic_evid}

Here we follow Ref. \cite{Dmitrasinovic:2014}.
First we show that the minimized action $S_{\rm min} = \int_{0}^{T} L(q(t), {\dot q}(t)) d t$ 
of a periodic orbit $q(t)$ in the strong potential $V(r)$, written as a time integral 
of twice the kinetic energy $T$ over period $T$, 
\begin{eqnarray}
S_{\rm min}(T) &=&  
2 \sum_{i=1}^3 \int_{0}^{T} \frac{{\bf p}_i^2}{2 m} d t 
= 2 \sum_{i=1}^3 \int_{{\bf r}_i(0)}^{{\bf r}_i(T)} {\bf p}_i \cdot d {\bf r}_i 
\end{eqnarray}
can be expressed as a closed-contour integral of a complex variable. 
After changing over to the relative-motion variables, one finds
\[S_{\rm min}(T) = 2 (\int_{{\bm \rho}(0)}^{{\bm \rho}(T)} {\bf p}_{\rho} \cdot d {\bm \rho}  
+ \int_{{\bm \lambda}(0)}^{{\bm \lambda}(T)} {\bf p}_{\lambda} \cdot d {\bm \lambda})\] 
The two independent three-body Jacobi two-vectors 
${\bm \rho}$ and ${\bm \lambda}$ can be replaced with two complex variables 
\[z_{\rho} = \rho_x + i \rho_y, z_{\lambda} = \lambda_x + i \lambda_y,\] 
so that the action $S_{\rm min}$, 
can be rewritten as a (double) closed contour integral in two complex variables: 
\[S_{\rm min}(T) = 2(\int_{z_{\rho(0)}}^{z_{\rho(T)}} {\dot z}_{\rho}^{*} d z_{\rho}  + 
\int_{z_{\lambda}(0)}^{z_{\lambda}(T)} {\dot z}_{\lambda}^{*} d z_{\lambda}).\]
Note that the periodicity of motion ${\bm \rho}(0) = {\bm \rho}(T)$, 
${\bm \lambda}(0) = {\bm \lambda}(T)$ implies $z_{\rho}(T) = z_{\rho}(0)$ and
$z_{\lambda}(T)=z_{\lambda}(0)$, which makes this integral a closed contour one
\[S_{\rm min} = 2 (\oint_{C_{\rho}} {\dot z}_{\rho}^{*} d z_{\rho}  + 
\oint_{C_{\lambda}} {\dot z}_{\lambda}^{*} d z_{\lambda})\] 
with its value determined by the residue theorem for functions of two complex variables. 

The existence and positions of 
poles in this (double) contour integral are not manifest in its present form; the same integral 
is given by $S_{\rm min} = - 2\int_{0}^{T} V(r(t)) d t$, 
due to the virial theorem, however, where the potential $V({r}(t))$ is 
known to have singularities on the shape sphere and the time-evolution dependence $r(t)$ of the periodic 
orbit parametrizes the contour. We remind the reader that the identity of two different forms of 
$S_{\rm min}$ 
holds only for periodic 
orbits, even though Cauchy's residue theorem holds more generally. 

Note the following implications of this formula:
1) for non-singular potentials ($\alpha < 0$) there are no poles in the potential, and consequently
no poles encircled by the contour, so the residue vanishes;
2) for singular potentials ($2 > \alpha > 0$) there are poles in the potential, but the residue 
depends 
on the integration contour, i.e., on the trajectory one the shape sphere and its topology $w$;
3) if the integration contour, i.e., the trajectory one the shape sphere repeats k-times the 
topologically equivalent path, then, for singular potentials ($2 > \alpha > 0$), the residue equals 
k times the single path residue. 

Next, we switch from the real $({\bm \rho},{\bm \lambda})$, or complex $(z_{\rho}, z_{\lambda})$ 
Cartesian Jacobi variables to the curvilinear hyper-spherical variables: the real hyper-radius $R$ and 
the overall rotation angle $\Phi = \frac12 (\varphi_{\rho} + \varphi_{\lambda})$, 
and the two angles parametrizing the shape-sphere, e.g. 
$(\theta = (\varphi_{\rho} - \varphi_{\lambda}),\chi = 2 {\rm Tan}^{-1}(\frac{\rho}{\lambda}))$.
Equivalently, we may use the complex variables $Z$, defined by $(R,\Phi)$ and $z$, defined by way 
of a stereographic projection from the shape-sphere parametrized by $(\theta,\chi)$. 
The ``variable'' $Z$ remains constant for many periodic orbits with zero angular momentum
in the strong potential ($\alpha = 2$), because $R$= const.. 
The condition $\Phi$ = const. is trickier, however, because there may be ``relatively periodic'' solutions 
with vanishing angular momentum ($L=0$) and a non-zero change $\Delta \Phi \neq 0$ of angle $\Phi$ 
over one period. All of the orbits
from the three sequences considered above have $\Delta \Phi$ = 0 over one period, however.
It is only for such orbits that we may eliminate the complex variable $Z$ from further consideration, 
and the problem becomes the (much) simpler one, of a single complex variable.

So, we see that the complex integration contour $C_{z}$ relevant to Cauchy's theorem, 
$S_{\rm min} = 2 i \pi \sum {\rm Res}$, for the above three sequences of periodic orbits 
is determined solely by the orbit's trajectory on the shape sphere: 
the only poles relevant to this contour integral are the two-body collision points on 
the shape sphere.
Consequently, the periodic orbits' minimized action (integral) is determined (only) by the 
topology of the closed contour on the shape sphere, i.e., by the homotopy group element 
of the periodic orbit. 
Repeated $k$-fold loops of the contour lead to $k$ times the initial integral, provided that no
branch cut is crossed in the process; otherwise the residue(s) at the pole(s) may change with 
the value of $k$. We have shown already in 
\ref{ss:Action_SCM} that each of the three 
poles is also a logarithmic branch cut, which implies a complicated structure of branch cuts
and (most probably) changing values of residues. Detailed study of analytic properties
of the action should be a subject of 
interest to mathematicians, \cite{Feynman:1964}.


\section*{References}
\begin{thebibliography}{9}

\bibitem{Jacobi1843} 
Jacobi C G J. 1969 
\textit{Vorlesungen {\" u}ber Dynamik} 
(New York, NY: Chelsea) (in German)

\bibitem{Poincare:1896}
Poincar\'e H. 1896 ``Sur les solutions p\'eriodiques et le principe de moindre action'', C.R.A.S. 123,
915-918 (in French), available at 
{\tt http://henripoincarepapers.univ-lorraine.fr/bibliohp/?}
{\tt \&a=on\&ln=Poincar\%C3\%A9\&action=go\&page=3}

\bibitem{Calogero:1969xj} 
  Calogero F. 1969
\textit{J.\ Math.\ Phys.\  10}, 2191

\bibitem{Khandekar:1972}
Khandekar D C, and Lawande S V. 1972 
\textit{Am. J. Phys.} {\bf 40}, 458

\bibitem{Moser1975}
Moser J. 1975 
\textit{Advances in Math.} {\bf 16} 197-220.

\bibitem{Poisson:2014}
Poisson E and Will C M. 2014 {\it Gravity Newtonian, Post-Newtonian, Relativistic}
(New York Cambridge University Press). 

\bibitem{Bloom:2014}
Bloom R S. 2014 ``Few-body collisions in a quantum gas mixture
$^{40}$K and $^{87}$Rb atoms'', PhD thesis,  Colorado U. Boulder (2014) (unpublished).

\bibitem{Barth:2015}
Barth M and 
Hofmann J. 2015  
\textit{Phys. Rev.} A {\bf 92} 062716

\bibitem{Moore1993}
Moore C. 1993 
\textit{Phys. Rev. Lett.} {\bf 70} 3675

\bibitem{Montgomery2004}
Montgomery R. 2005 
\textit{Ergod. Th. \& Dynam. Sys.} {\bf 25} 921-947 
\textit{ArXiv: math/0405014v1 [math.DS]} 

\bibitem{Fujiwara2004a}
Fujiwara T, Fukuda H, Kameyama A, Ozaki H and Yamada M. 2004 
\textit{J. Phys.} A {\bf 37}, 10571

\bibitem{Montgomery2002}
Montgomery R. 2002 
\textit{Arch. Rat. Mech. Anal.} {\bf 164} 311-340.

\bibitem{Tosel2000}
Tosel E J. 2000 
Arch. Rat Mech. Anal. {\bf 152}, 187-205

\bibitem{Suvakov:2013}
\v Suvakov M. and Dmitra\v sinovi\' c V. 2013
\textit{Phys.\ Rev.\ Lett.} {\bf 110} 114301

\bibitem{Suvakov:2013b}
\v Suvakov M. 2014:
\textit{Celest. Mech. Dyn. Astron.} {\bf 119} 369-377 
 
\bibitem{Shibayama:2015}
\v Suvakov M and Shibayama M. 2016: 
\textit{Celest. Mech. Dyn. Astron.} {\bf 124} 155-162 

\bibitem{Martynova2009}
Martynova A I, Orlov V V and Rubinov A V. 2009 
\textit{Astron. Rep.} {\bf 53} 710

\bibitem{Iasko2014}
Iasko P P and Orlov  V V. 2014
\textit{Astron. Rep.} {\bf 58} 869-879

\bibitem{Suvakov:2014}
\v Suvakov M., Dmitra\v sinovi\' c V. 2014
\textit{Am. J. Phys.} {\bf 82} 609-619.

\bibitem{Dmitrasinovic:2016}
Dmitra\v{s}inovi\'{c} V, Hudomal A., Shibayama M and Sugita A. 2017
{\it ``Newtonian Periodic Three-Body Orbits with Zero Angular Momentum: 
Linear Stability and Topological Dependence of the Period''}
arXiv:1705.03728 [physics.class-ph]

\bibitem{Jankovic:2015}
Jankovi{\' c} M R and Dmitra\v sinovi{\' c} V. 2016:  
\textit{Phys. Rev. Lett.} {\bf 116} 064301

\bibitem{Dmitrasinovic:2015}
Dmitra\v{s}inovi\'{c} V and \v{S}uvakov M. 2015 
\textit{Phys. Lett.} A {\bf 379}, 1939-1945
 
\bibitem{Landau}
Landau L D and Lifshitz E M. 1976 \textit{Mechanics},
(Oxford, UK: Butterworth-Heinemann) (3rd ed.) Sec.~10.

\bibitem{Suvakov:2010}
\v Suvakov M and ~Dmitra\v sinovi\' c V. 2011
\textit{Phys. Rev.} E {\bf 83}, 056603

\bibitem{Hairer1993}
Hairer E, 
N{\o}rsett S and 
Wanner G. 1993  
\textit{Solving Ordinary Differential Equations I: Nonstiff Problems}, 2nd ed. 
(Springer-Verlag, Berlin) 

\bibitem{Python}
Python open-source library for mathematics, science, and engineering: 
{\tt http://www.scipy.org}

\bibitem{Montgomery1998}
Montgomery R. 1998 
\textit{Nonlinearity} {\bf 11}, 363 - 376

\bibitem{Pars:1964}
Pars L A. 1965: \textit{Analytical Dynamics} (London, UK: Heinemann). 

\bibitem{Siegel}
Siegel C L and Moser J K. 1971 \textit{Lectures on Celestial mechanics},  
(Berlin, Germany: Springer).

\bibitem{Sundman:1907}
Sundman K F. 1907  
\textit{Acta Soc. Sci. Fennicae {\bf 34}}, no. 6, 144-151; 
Sundman K F. 1912 \textit{Acta Math.} {\bf 36} 105-179 (in French)

\bibitem{Henkel:2001}
Henkel M. 2001 
\textit{Philosophia Scientiae} {\bf 5 (2)} 161-184 
  
\bibitem{Dmitrasinovic:2014}
Dmitra\v sinovi\' c V. 2014 
\textit{Topological dependence of Kepler's third law for three-body orbits}, in
the Proceedings of the Chiba \textit{Symposium on Celestial Mechanics and N-body Dynamics 2014}, see

\bibitem{Hakobyan:2009ac} 
  Hakobyan T,~Krivonos S,~Lechtenfeld O and~Nersessian A. 2010
\textit{Phys.\ Lett.} A {\bf 374} 801

\bibitem{Hakobyan:2010ia} 
  Hakobyan T,~Lechtenfeld O, Nersessian A and Saghatelian A. 2011
\textit{J.\ Phys.} A {\bf 44} 055205 

\bibitem{Calogero2001}
Calogero F. 2001 
\textit{Classical Many-Body Problems Amenable to Exact Treatments},
(Berlin, Germany: Springer).  

\bibitem{Feynman:1964}
Goodstein D L and Goodstein J R. 1999 \textit{Feynman's Lost Lecture: The Motion of Planets 
Around the Sun} (New York, NY: W - W. Norton \& company)

\end{thebibliography}
\end{document}